\documentclass[11pt]{article}
\usepackage[margin=1in,headheight=14pt,headsep=0.2in]{geometry}
\usepackage{times}
\usepackage{natbib}
\usepackage{fancyhdr}
\setcitestyle{authoryear,round,citesep={;},aysep={,},yysep={;}}
\usepackage{amsmath,amsfonts,bm}

\def\eqref#1{equation~\ref{#1}}

\def\1{\bm{1}}

\DeclareMathAlphabet{\mathsfit}{\encodingdefault}{\sfdefault}{m}{sl}
\SetMathAlphabet{\mathsfit}{bold}{\encodingdefault}{\sfdefault}{bx}{n}

\usepackage{url}
\usepackage{graphicx}
\usepackage[percent]{overpic}
\usepackage{array}
\usepackage{booktabs}
\usepackage{algorithm}
\usepackage{algorithmic}
\usepackage{makecell}
\usepackage{multirow}
\usepackage{listings}
\usepackage{longtable}
\usepackage[colorlinks=true,linkcolor=red,citecolor=blue,urlcolor=magenta]{hyperref}

\renewcommand{\headrulewidth}{0.5pt}
\fancypagestyle{plain}{%
  \fancyhf{}
  \fancyfoot[C]{\thepage}
  \renewcommand{\headrulewidth}{0.5pt}
}

\lstdefinestyle{prompt}{
  basicstyle=\ttfamily\scriptsize,
  columns=fullflexible,
  keepspaces=true,
  showstringspaces=false,
  breaklines=true,
  breakatwhitespace=false,
  frame=single,
  framerule=0.3pt,
  xleftmargin=0.4em,
  xrightmargin=0.4em,
  aboveskip=4pt,
  belowskip=6pt
}

\newcommand{\bestresult}[1]{\textbf{#1}}
\newcommand{\secondresult}[1]{\underline{#1}}

\title{CosmosAlpha: Multi-Relational Graph Search with Pareto-Guided Exploration for LLM-Driven Alpha Mining}

\author{
\href{mailto:wangzhanhe19@mails.ucas.ac.cn}{\textcolor{black}{Zhan-He Wang}}\textsuperscript{1}\quad
\href{mailto:langziyang413@sjtu.edu.cn}{\textcolor{black}{Zi-Yang Lang}}\textsuperscript{2}\quad
\href{mailto:liyiyang20@mails.ucas.ac.cn}{\textcolor{black}{Yi-Yang Li}}\textsuperscript{1}\quad
\href{mailto:lanqingyu19@mails.ucas.ac.cn}{\textcolor{black}{Qing-Yu Lan}}\textsuperscript{1}\quad
\href{mailto:yspiao@ucas.edu.cn}{\textcolor{black}{Yun-Song Piao}}\textsuperscript{1,3}\\[0.5em]
{\small \textsuperscript{1}Artificial Intelligence and Quantitative Finance Laboratory,}\\
{\small School of Physical Sciences, University of Chinese Academy of Sciences}\\
{\small \textsuperscript{2}School of Physics and Astronomy, Shanghai Jiao Tong University}\\
{\small \textsuperscript{3}School of Fundamental Physics and Mathematical Sciences,}\\
{\small Hangzhou Institute for Advanced Study, UCAS}\\[0.5em]
{\footnotesize \texttt{wangzhanhe19@mails.ucas.ac.cn}, \texttt{langziyang413@sjtu.edu.cn}}\\
{\footnotesize \texttt{liyiyang20@mails.ucas.ac.cn}, \texttt{lanqingyu19@mails.ucas.ac.cn}, \texttt{yspiao@ucas.edu.cn}}
}
\date{}

\begin{document}

\maketitle

\begin{abstract}
Discovering predictive alpha factors from noisy financial data is a fundamental yet challenging
problem in quantitative finance. LLM-driven formulaic alpha mining offers an attractive paradigm
by combining interpretable symbolic expressions with iterative generation and refinement guided by quantitative feedback.
However, existing generation-centric and tree-based search strategies primarily organize factors through
generation lineage, leaving structural and statistical relationships across search branches
underutilized. Consequently, information remains localized along parent--child trajectories,
which can confine exploration to redundant regions.
Moreover, effective factor discovery must account for multiple distinct objectives that cannot be
adequately captured by a single search criterion.
We introduce \textbf{CosmosAlpha}, a graph-based LLM factor-mining framework that represents
discovered factors as a multi-relational graph encoding generation lineage, expression-structure
similarity, and cross-sectional exposure correlation. A Graph Flow mechanism propagates search
utilities across the graph, allowing each factor to be assessed using relational evidence from nearby
factors rather than in isolation. Pareto-guided selection jointly considers quality, fertility and diversity, explicitly modeling their trade-offs within a multi-objective search
framework. Across CSI 300 and CSI 500, CosmosAlpha-Graph achieves the best
mean rank over seven predictive and portfolio metrics and outperforms its
controlled CosmosAlpha-MCTS counterpart on every reported metric, showing a
consistent advantage for relational graph search in LLM-driven factor mining.
The source code is available at \url{https://github.com/wzhmtldswz/CosmosAlpha}.

\end{abstract}

\section{Introduction}
\label{sec:introduction}

Alpha mining seeks to discover quantitative signals from market data for
predicting future stock returns~\citep{jensen1968performance,famamacbeth1973risk,gu2020empirical}. 
In modern quantitative research, an \emph{alpha
factor} is a predictive signal used to rank stocks by their expected future
returns; its predictive value is commonly assessed by the alignment between factor
scores and subsequent cross-sectional returns and by the performance of
portfolios formed from those scores
~\citep{fama1992crosssection,gu2020empirical}. In formulaic
alpha mining, the factor is represented as an explicit mathematical expression
over observable market variables, making its construction and evaluation
transparent~\citep{kakushadze2016formulaic,yu2023alphagen}. Such expressions can
capture diverse patterns in price, volume, and other market data. Established
handcrafted libraries such as Alpha158 provide fixed collections of interpretable
factors and serve as reference points for automated mining~\citep{yang2020qlib}.
Individual factors, whether handcrafted or automatically mined, are typically
weak and noisy. Reliable discovery therefore requires searching a large
expression space while controlling spurious correlations and redundant
candidates~\citep{harvey2016crosssection,gu2020empirical}.

\begin{figure}[!t]
\centering
\includegraphics[width=\textwidth]{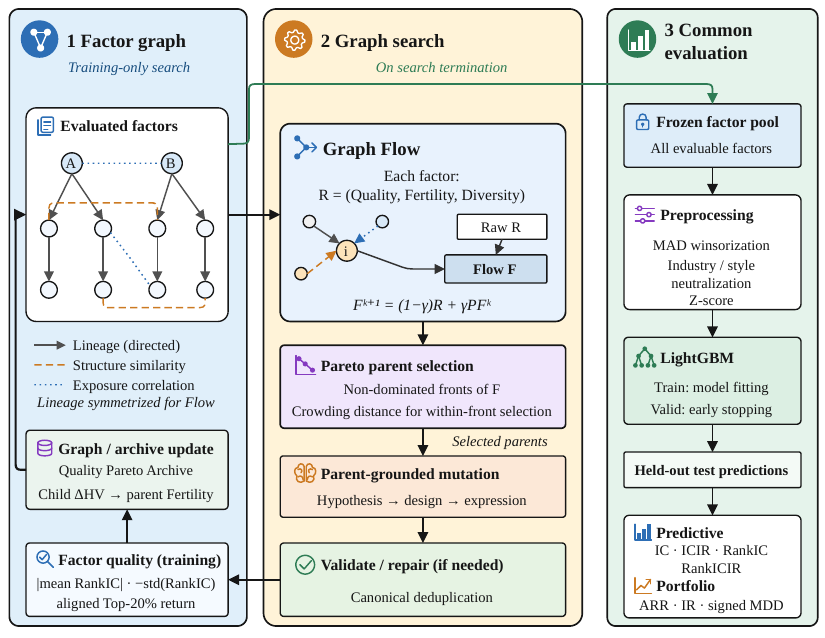}
\caption{Overview of CosmosAlpha. (1) Evaluated factors form a multi-relational
graph with generation lineage, structure similarity, and exposure correlation.
(2) Graph Flow propagates Quality, Fertility, and Diversity; Pareto fronts and
crowding distance select parents for LLM-driven mutation. Valid unique children
are evaluated on training data and update the graph, Quality Pareto Archive,
and parent Fertility. (3) After search terminates, the full evaluable factor
pool is frozen for common preprocessing, LightGBM integration, and held-out
predictive and portfolio evaluation.}
\label{fig:cosmosalpha-overview}
\end{figure}

Automated alpha mining has progressively expanded both the symbolic search space
and the feedback used to navigate it. Evolutionary approaches, including genetic
programming (GP), search formula structures through mutation and selection;
AutoAlpha and AlphaEvolve are representative examples of this direction
~\citep{zhang2020autoalpha,cui2021alphaevolve}, while reinforcement-learning
methods optimize expression construction or the quality of an emerging factor
set~\citep{yu2023alphagen,zhao2025qfr,zhao2025tlrs,zhu2025alphaqcm}. Large
language models provide another route by introducing financial hypotheses,
semantic reasoning, and iterative refinement from quantitative feedback
~\citep{li2024fama,yu2026hybridllm,tang2025alphaagent,luo2026alphabench,
liu2026cognitive}.
Among recent LLM-based approaches, several systems also
retain richer persistent search states: Alpha Jungle organizes refinements with
Monte Carlo Tree Search (MCTS)~\citep{shi2026alphajungle}, AlphaPROBE navigates
a lineage-based directed acyclic graph (DAG)~\citep{guo2026alphaprobe},
QuantaAlpha evolves complete mining trajectories~\citep{han2026quantaalpha},
AlphaAgentEvo learns multi-turn evolution through agentic reinforcement learning
~\citep{tang2026alphaagentevo}, and X-Alpha guides research across cycles with
persistent memory~\citep{liu2026xalpha}. Together, these developments move
formulaic alpha mining from isolated generation toward iterative search that
can reuse previous evaluations.

Despite this progress, many recent LLM-based methods organize persistent search
state around provenance, trajectories, or research memory. A generation edge
records \emph{who generated whom}, but it does not reveal whether factors from
otherwise distant search paths reuse similar subexpressions or induce nearly
identical exposure profiles across stocks. LLM-based systems address parts of
this problem in different ways: AlphaAgent regularizes originality at the
abstract syntax tree (AST) level, Alpha Jungle suppresses frequent subtrees, and
AlphaPROBE reasons over global lineage
~\citep{tang2025alphaagent,shi2026alphajungle,guo2026alphaprobe}. In a
complementary GFlowNet-based approach, AlphaSAGE uses a structure-aware encoder
for each expression's AST while promoting diverse generation
~\citep{alphasage2026}. These approaches expose complementary signals, but they
do not jointly treat generation history, structure similarity, and empirical
cross-sectional exposure correlation as coexisting relation types between
\emph{complete evaluated factors}. Such factor-to-factor relations could help
reuse evidence across otherwise distant search paths and reduce redundant
exploration.

Selecting which factors to expand remains a difficult part of automated alpha
mining. We argue that a candidate parent should be assessed along three complementary
dimensions: current quality, fertility, and diversity. Current \emph{quality}
measures a factor's present predictive value; \emph{fertility} measures whether its
previous expansions produced high-quality descendants; and \emph{diversity} measures
whether it occupies a less redundant region of the factor pool. Existing methods often
either prioritize quality alone or reduce these criteria to a fixed scalar,
which can obscure meaningful trade-offs among candidate parents.
Structure similarity and cross-sectional exposure
correlation provide direct measures of redundancy for diversity estimation,
while generation links preserve search provenance for relational propagation.
Pareto-based multi-objective selection can then retain non-dominated trade-offs
without collapsing these criteria to a single scalar
~\citep{zitzler1999spea,deb2002nsga2}.

We introduce \textbf{CosmosAlpha}, an LLM-driven framework that treats the
evaluated factor pool as an active multi-relational search state
(Figure~\ref{fig:cosmosalpha-overview}). Each node is a complete factor, and
three relation types capture generation lineage, expression structure, and
cross-sectional exposure correlation among factors. Graph Flow propagates node
utilities through these typed neighborhoods so that parent assessment reflects
both local measurements and evidence from related factors. Pareto-guided
selection then preserves trade-offs among quality, fertility, and diversity,
and the selected factors serve as parents for subsequent LLM-guided mutations.
After evaluation, valid children enter the graph and update the state used by
the next round. The factor archive is thus no longer a passive record of past
attempts; it becomes the state that organizes subsequent exploration.

We evaluate CosmosAlpha on CSI 300 and CSI 500 under a common data and downstream
LightGBM integration protocol. CosmosAlpha-Graph achieves the best overall mean
rank on both universes and outperforms the controlled CosmosAlpha-MCTS
counterpart on every reported predictive and portfolio metric. Component and
relation ablations support the contributions of relational propagation and
multi-objective selection, while experiments across three LLM backbones and
three independent runs show that the overall advantage is not confined to a
single model or stochastic run.

Our contributions are summarized as follows:
\begin{itemize}
    \item We formulate LLM-driven formulaic alpha mining as iterative expansion
    over a \emph{pool-level multi-relational graph} that jointly represents
    generation lineage, expression structure, and empirical exposure
    correlation between complete factors.

    \item We develop Graph Flow and Pareto-guided expansion to combine
    relational evidence with distinct quality, fertility, and diversity
    criteria, preserving useful search trade-offs without fixed scalarization.

    \item We provide a controlled evaluation in which CosmosAlpha-Graph and
    CosmosAlpha-MCTS share the factor language, evaluator, LLM backbone, and
    downstream modeling protocol, with comparable factor-pool sizes in the main
    experiments and a separate comparison under equal Token budgets;
    cross-universe comparisons, ablations, LLM sensitivity, and multi-run analysis
    further test the robustness of the design.
\end{itemize}

\section{Related Work}
\label{sec:related-work}

\paragraph{Formulaic alpha mining.}
Automated formulaic alpha mining has progressed from evolutionary search to
methods that optimize factor sets and use learned feedback. Genetic programming
systems such as AutoAlpha use hierarchical evolution and quality-diversity
guidance~\citep{zhang2020autoalpha}, while AlphaEvolve expands formula
representations and prunes redundant candidates~\citep{cui2021alphaevolve}.
Reinforcement-learning and feedback-driven generative methods instead explore
factor-set construction and richer feedback mechanisms. AlphaGen and AlphaForge
optimize factor-set construction, while QuantFactor REINFORCE and
trajectory-level reward shaping improve the stability and density of RL signals;
RiskMiner and AlphaQCM explore alternative tree-search and distributional-RL
objectives~\citep{yu2023alphagen,shi2025alphaforge,zhao2025qfr,zhao2025tlrs,
ren2024riskminer,zhu2025alphaqcm}. These approaches broaden formula generation
and factor-set construction, but generally do not represent the complete
evaluated factor pool as a state with explicit relations among its members.

\paragraph{LLM-driven and agentic alpha search.}
LLM-based systems add semantic hypothesis generation and feedback-driven
refinement to formula mining. FAMA and recent hybrid LLM--RL frameworks integrate
LLM proposals with symbolic or optimization loops, while AlphaAgent organizes
generation and feedback through a multi-agent workflow
~\citep{li2024fama,yu2026hybridllm,tang2025alphaagent}.
CogAlpha combines LLM reasoning with code-level mutation and crossover guided
by financial feedback~\citep{liu2026cognitive}.
AlphaBench provides a systematic evaluation suite across factor generation,
evaluation, and search~\citep{luo2026alphabench}. More structured systems
preserve search history explicitly: Alpha Jungle uses MCTS, QuantaAlpha evolves
complete trajectories, and AlphaAgentEvo learns multi-turn evolution with
agentic reinforcement learning
~\citep{kocsis2006uct,shi2026alphajungle,han2026quantaalpha,
tang2026alphaagentevo}. R\&D-Agent-Quant extends the scope to coordinated factor
and model R\&D~\citep{li2025rdagentquant}. These methods retain richer workflow
state, but it remains tied to a particular trajectory or task-specific agent
state rather than a shared factor-level representation. The remaining question
is how the pool itself can serve as persistent relational state for formula
search.

\paragraph{Relational graphs and multi-objective search.}
The closest structural precedents represent either factor lineage or an
individual expression. AlphaPROBE represents complete factors in a lineage-based
DAG for retrieval and biased evolution~\citep{guo2026alphaprobe}; AlphaSAGE
encodes the AST of an individual expression with a structure-aware graph encoder
and uses GFlowNets to promote diverse high-reward generation
~\citep{bengio2021gflownet,alphasage2026}. Graph Flow is conceptually related to
diffusion-style propagation over graph neighborhoods, which shares information
among locally connected nodes~\citep{zhou2003localglobal,klicpera2019appnp}.
CosmosAlpha applies this principle over complete evaluated factors, with distinct
relations for generation provenance, structure similarity, and cross-sectional
exposure correlation, rather than learning an expression encoder or a generative
flow policy. This factor-pool view jointly exposes relation types that prior
methods handle separately, enabling the same state to support both information
propagation and parent selection.

Multi-objective evolutionary optimization provides the complementary selection
principle. Pareto archives and diversity-preserving selection maintain solutions
with different trade-offs instead of imposing a single weighted objective
~\citep{zitzler1999spea,deb2002nsga2}. CosmosAlpha applies this principle to
propagated quality, fertility, and diversity, separating graph-based evidence
sharing from parent selection and preserving non-dominated trade-offs during
parent selection.

\section{Method}
\label{sec:method}

\subsection{Problem Formulation and Overview}
\label{sec:method-overview}

We consider formulaic alpha discovery over a time-varying stock universe. Let
$\mathcal{X}_t\in\mathbb{R}^{N_t\times d}$ denote the market observations
available on day $t$, and let a formulaic factor $f_i$ produce a cross-sectional
exposure vector $\mathbf z_{i,t}=f_i(\mathcal{X}_{\leq t})$. Given a finite
search budget, the objective is to discover a factor library that is predictive
while avoiding repeated exploration of redundant regions.

Existing LLM-based alpha-mining methods commonly organize search as sequential
refinement, tree search, or generation-centric evolution
\citep{tang2025alphaagent,shi2026alphajungle,guo2026alphaprobe}. These
representations primarily preserve \emph{who generated whom}, but factors from
different search branches may still share similar expressions or induce highly
correlated stock exposures. CosmosAlpha therefore models factor discovery as
search over a multi-relational graph
\begin{equation}
\mathcal{G}=(\mathcal{V},\mathcal{E}^{\mathrm{gen}},
\mathcal{E}^{\mathrm{str}},\mathcal{E}^{\mathrm{corr}}),
\label{eq:factor-graph}
\end{equation}
where nodes are evaluated factors and the three edge types encode generation
lineage, expression structure, and empirical exposure correlation. Each node is
scored along quality, fertility, and diversity; these utilities are
propagated through the graph by Graph Flow, and Pareto-guided selection chooses
parents for LLM mutation. Valid children are evaluated and inserted back into
the graph, yielding an iterative graph-search process.

\subsection{Multi-Relational Factor Graph}
\label{sec:multi-relational-graph}

The three relations capture complementary notions of factor proximity.

\paragraph{Generation relation.}
If factor $f_j$ is produced by mutating factor $f_i$, we record a directed
provenance edge $i\rightarrow j$. For message propagation we use the
symmetrized direct parent--child adjacency
\begin{equation}
W^{\mathrm{gen}}_{ij}
=\mathbb{I}[i\text{ and }j\text{ are a direct parent--child pair}].
\label{eq:generation-adj}
\end{equation}
Longer-range lineage information is captured by multi-step propagation rather
than explicit ancestor--descendant edges.

\paragraph{Structure relation.}
Let $\mathcal{S}_i$ be the set of canonical subexpressions extracted from the
expression tree of $f_i$. We define
\begin{equation}
S^{\mathrm{str}}_{ij}
=\frac{|\mathcal{S}_i\cap\mathcal{S}_j|}
{|\mathcal{S}_i\cup\mathcal{S}_j|},
\qquad
d^{\mathrm{str}}_{ij}=1-S^{\mathrm{str}}_{ij}.
\label{eq:structure-similarity}
\end{equation}
Canonicalization preserves operator, topology, field, and parameter identities
while removing spurious differences caused by commutative argument ordering.

\paragraph{Correlation relation.}
Expression structure does not fully determine empirical factor behavior. We
therefore measure behavioral proximity by first computing the daily
cross-sectional Pearson correlation and then averaging equally over valid dates:
\begin{equation}
C_{ij}=\frac{1}{|\mathcal{T}_{ij}|}
\sum_{t\in\mathcal{T}_{ij}}
\operatorname{Pearson}_{\mathrm{cs}}
(\mathbf z_{i,t},\mathbf z_{j,t}),
\qquad
d^{\mathrm{corr}}_{ij}=1-|C_{ij}|.
\label{eq:factor-corr}
\end{equation}
The absolute value treats strongly negative and strongly positive exposure
correlations as equally redundant up to factor orientation. The structure and
correlation graphs retain only local neighbors; exact construction rules are
reported in Appendix~\ref{app:graph-construction}.

Let $W^m$ denote the weighted adjacency matrix for relation
$m\in\{\mathrm{gen},\mathrm{str},\mathrm{corr}\}$. For the retained structure
and correlation neighbors, the edge weights are given by the corresponding
similarities: $W^{\mathrm{str}}_{ij}=S^{\mathrm{str}}_{ij}$ and
$W^{\mathrm{corr}}_{ij}=|C_{ij}|$. Each modality is row-normalized as
$P^m=(D^m)^{-1}W^m$, where
$D^m_{ii}=\sum_j W^m_{ij}$.
We combine the three modalities as
\begin{equation}
\boxed{
P=\frac{1}{3}
\left(P^{\mathrm{gen}}+P^{\mathrm{str}}+P^{\mathrm{corr}}\right).
}
\label{eq:multi-relation-transition}
\end{equation}
Separate normalization prevents denser similarity graphs from overwhelming the
sparser generation graph.

\subsection{Multi-Objective Search Utility}
\label{sec:node-utility}

A useful parent should not be judged only by its current predictive strength:
it may also be valuable because it repeatedly generates strong descendants or
because it occupies an underexplored region. CosmosAlpha therefore assigns
three complementary utilities to each node.

\paragraph{Intrinsic quality.}
We characterize standalone factor quality by
\begin{equation}
\mathbf q_i=
\left(
|\overline{\mathrm{RankIC}}_i|,
-\sigma(\mathrm{RankIC}_i),
R^{\mathrm{Top20}}_i
\right),
\label{eq:quality-vector}
\end{equation}
combining predictive strength, temporal stability, and a direction-aligned
Top-20\% return metric. A \emph{Quality Pareto Archive} retains the
non-dominated factors under these raw objectives. For scalar node quality
and hypervolume computation, the three objectives are normalized using fixed
anchors and clipped to $[0,1]$. The normalized coordinates are averaged to
obtain $Q_i^{\mathrm{self}}$ (Appendix~\ref{app:quality-details}).

\paragraph{Fertility.}
Current quality does not measure whether a node reliably produces useful
descendants. Let $\mathcal{A}^{-}_r$ denote the Quality Pareto
Archive frozen before round $r$. For a newly evaluated child $c$, its birth-time
quality contribution is
\begin{equation}
\boxed{
\Delta HV_c=
HV(\mathcal{A}^{-}_r\cup\{c\})-HV(\mathcal{A}^{-}_r).
}
\label{eq:delta-hv}
\end{equation}
All children in the same round use the same frozen archive, avoiding
order-dependent rewards. The fertility of parent $i$, denoted
$Q_i^{\mathrm{fert}}$, aggregates its best historical expansion outcomes, so it
measures the node's demonstrated ability to produce frontier-expanding
descendants rather than its current quality alone. The exact attempt-level
aggregation is specified in Appendix~\ref{app:fertility-details}.

\paragraph{Exploration diversity.}
At the beginning of round $r$, we select a high-quality reference set
$\mathcal{R}_r$ from the current Quality Pareto Archive and keep it fixed while
computing diversity for that round. To discourage concentration around a few
successful factor families, we measure novelty relative to this set.
Using the structure and correlation distances defined above,
\begin{equation}
Q_i^{\mathrm{div}}
=\frac{1}{2}\left[
\operatorname{AvgKNN}_{j\in\mathcal{R}_r}d^{\mathrm{str}}_{ij}
+
\operatorname{AvgKNN}_{j\in\mathcal{R}_r}d^{\mathrm{corr}}_{ij}
\right].
\label{eq:average-distance}
\end{equation}
Generation distance is excluded to avoid double-counting lineage information
already represented in the graph topology. Reference-set construction is
reported in Appendix~\ref{app:diversity-reference}.

The resulting node utility is the three-dimensional vector
\begin{equation}
\boxed{
\mathbf R_i=(Q_i^{\mathrm{self}},Q_i^{\mathrm{fert}},Q_i^{\mathrm{div}}).
}
\label{eq:node-reward}
\end{equation}
We deliberately preserve these objectives instead of assigning a fixed scalar
trade-off among current quality, fertility, and diversity.

\subsection{Graph Flow and Pareto-Guided Expansion}
\label{sec:graph-flow}

The raw utility $\mathbf R_i$ reflects node-specific evidence but ignores the
quality of its neighborhood. The transition matrix $P$ in
Eq.~\ref{eq:multi-relation-transition} represents the generation, structure,
and correlation links between factors and sets the weights for aggregating
neighboring utilities. Stacking node utilities into $R$, Graph Flow iterates
\begin{equation}
\boxed{
F^{(k+1)}=(1-\gamma)R+\gamma P F^{(k)},
\qquad F^{(0)}=R,
}
\label{eq:graph-flow}
\end{equation}
where $\gamma\in[0,1)$ controls the contribution of neighboring nodes. The first
term preserves node-specific evidence, while the second propagates information
through generation, structure, and correlation neighborhoods. Setting
$\gamma=0$ removes graph propagation and yields a natural ablation.

Parent selection operates directly on the propagated vectors
$\mathbf F_i=(F_i^{\mathrm{self}},F_i^{\mathrm{fert}},F_i^{\mathrm{div}})$.
We apply non-dominated sorting rather than a manually weighted scalarization;
parents are drawn from successive Pareto fronts, and crowding distance resolves
a front that exceeds the remaining parent budget. This \emph{search Pareto}
mechanism is distinct from the intrinsic Quality Pareto Archive: the former
controls exploration in propagated utility space, whereas the latter measures
factor-quality frontier improvement for $\Delta HV$.

For each selected parent, the LLM receives its factor expression and search
context and generates parent-grounded mutations. Valid unique children are
evaluated, inserted into the graph, and used to update the Quality Pareto
Archive, parent fertility, and all affected graph relations. The complete round
semantics, budget rules, and pseudocode are provided in
Appendix~\ref{app:search-details}.

\section{Experiments}
\label{sec:experiments}

\subsection{Experimental Setup}
\label{sec:setup}

\paragraph{Datasets.}
We evaluate factor mining on the CSI 300 and CSI 500 universes using daily
A-share market data. Historical constituent membership and trading-eligibility
masks are applied point-in-time, and all methods use the same data snapshot and
preprocessing conventions. We use three chronological, non-overlapping
segments. The training set (April 1, 2010--May 31, 2019) provides feedback for
factor mining and is subsequently used to fit the downstream LightGBM
model~\citep{ke2017lightgbm}.
The validation set (June 1, 2019--December 31, 2020) is used only for LightGBM
early stopping. The test set (January 1, 2021--December 31, 2025) is held out
until the factor pools and fitted model configurations are fixed, and is used
only to compute final out-of-sample predictions and portfolio results. Neither
the validation nor test set is used to select factor expressions or tune the
search methods.

\paragraph{Baselines and mining protocol.}
We compare against the handcrafted Alpha158 factor library~\citep{yang2020qlib},
genetic programming (GP)~\citep{koza1992genetic}, and representative recent
automated factor-mining methods. The LLM-driven baselines include
AlphaAgent~\citep{tang2025alphaagent}, R\&D-Agent~\citep{li2025rdagentquant},
AlphaPROBE~\citep{guo2026alphaprobe}, and QuantaAlpha~\citep{han2026quantaalpha};
we additionally evaluate AlphaGen~\citep{yu2023alphagen}, AlphaForge
~\citep{shi2025alphaforge}, and AlphaSAGE~\citep{alphasage2026} as recent
factor-collection and structure-aware search methods.
We further include CosmosAlpha-MCTS as a controlled lineage-based search
baseline using the same factor language and evaluation stack as our graph
variant. AlphaJungle~\citep{shi2026alphajungle} is not included as a separate
row because we did not identify an official public implementation at the time
of submission; CosmosAlpha-MCTS therefore serves as a transparent MCTS-based
proxy. It maintains multiple subtrees rather than
a single large tree, which encourages more diverse factor-generation
trajectories while preserving the lineage-based search structure. Unless
otherwise specified, DeepSeek-V4-Flash is used as the LLM
backbone for LLM-driven methods. We target roughly 150 evaluable factors per
automated method, balancing mining cost and model input size near Alpha158.
A separate 4M-Token comparison controls search cost
(Appendix~\ref{app:matched-budget}). Each method retains its own search procedure; the
factors in its resulting pool are then recomputed on the common data snapshot
and evaluated with the shared preprocessing, LightGBM, and portfolio protocol.
Appendix~\ref{app:baselines}
describes the baselines and lists available public implementation repositories.

\paragraph{Evaluation protocol and metrics.}
To measure the information contained in a discovered factor \emph{set}, we
recompute each method's factor pool under the common data protocol and use the
factors jointly as LightGBM features. The same preprocessing, model, and
backtest configuration is used for every method; metric definitions and
implementation details are given in Appendices~\ref{app:metrics}
and~\ref{app:experiment-details}, respectively. Predictive performance is
measured from held-out LightGBM predictions using IC, ICIR, RankIC, and
RankICIR. Portfolio performance is measured from the same signal using
annualized return (ARR), information ratio (IR), and signed maximum drawdown
(MDD). Higher is better for the first six metrics; because MDD is reported as a
negative return, a value closer to zero is better.

\subsection{Main Results}
\label{sec:main-results}

Tables~\ref{tab:main-csi300} and~\ref{tab:main-csi500} compare the integrated
factor pools on CSI 300 and CSI 500, respectively. On CSI 300,
CosmosAlpha-Graph attains the best mean rank (1.43), leading IC, RankIC,
RankICIR, ARR, and IR. AlphaPROBE leads ICIR, while
R\&D-Agent has the best signed MDD (-24.87\%) and CosmosAlpha-Graph is second
(-25.07\%). Alpha158 and AlphaForge tie for the next-best mean rank (3.86).
CosmosAlpha-Graph exceeds CosmosAlpha-MCTS on all seven metrics under the
shared evaluation protocol. The matched-generation and Token-budget control
in Appendix~\ref{app:matched-budget} further supports this controller-level
advantage.

On CSI 500, CosmosAlpha-Graph attains the best mean rank (1.29) and leads IC,
ICIR, RankIC, ARR, and signed MDD, while AlphaAgent leads IR. Alpha158 remains
strongest on RankICIR. CosmosAlpha-MCTS is the best non-Graph method by mean
rank (4.14) and is also the strongest non-Graph method on IC (0.0356), followed
by Alpha158 at 4.43. These results indicate that the graph variant's advantage
also extends to CSI 500, although Alpha158 retains an advantage on RankICIR.

\begin{table}[t]
\caption{CSI 300 test performance after common LightGBM integration. Graph and MCTS denote the CosmosAlpha variants. ARR and signed MDD are percentages. Mean rank is a descriptive average over seven metrics (rank 1 is best; ties receive average ranks). Bold and underline indicate the best and second-best values.}
\label{tab:main-csi300}
\centering
\small
\setlength{\tabcolsep}{4pt}
\renewcommand{\arraystretch}{1.15}
\begin{tabular*}{\linewidth}{@{\extracolsep{\fill}}lrrrrrrrr@{}}
\toprule
\multirow{2}{*}{Method} & \multicolumn{4}{c}{Predictive} & \multicolumn{3}{c}{Portfolio} & \multirow{2}{*}{Mean rank} \\
\cmidrule(lr){2-5} \cmidrule(lr){6-8}
& IC & ICIR & RankIC & RankICIR & ARR & IR & MDD & \\
\midrule
Alpha158 & \secondresult{0.0320} & \secondresult{0.3474} & \secondresult{0.0479} & \secondresult{0.5668} & 6.32 & 0.3945 & -29.73 & \secondresult{3.86} \\
GP & 0.0231 & 0.3098 & 0.0300 & 0.4483 & 4.32 & 0.2215 & -42.45 & 9.71 \\
AlphaGen & 0.0301 & 0.3128 & 0.0452 & 0.5461 & 4.25 & 0.2490 & -32.47 & 7.29 \\
AlphaForge & 0.0304 & 0.3404 & 0.0469 & 0.5615 & 6.88 & 0.4164 & -25.41 & \secondresult{3.86} \\
AlphaSAGE & 0.0243 & 0.2574 & 0.0432 & 0.4750 & 6.57 & 0.4007 & -27.50 & 6.57 \\
AlphaAgent & 0.0235 & 0.2724 & 0.0390 & 0.4753 & 7.13 & \secondresult{0.4393} & -26.54 & 6.00 \\
R\&D-Agent & 0.0173 & 0.1811 & 0.0356 & 0.3945 & 5.99 & 0.3667 & \bestresult{-24.87} & 8.43 \\
QuantaAlpha & 0.0211 & 0.2154 & 0.0393 & 0.4167 & \secondresult{7.22} & 0.4303 & -26.11 & 6.71 \\
AlphaPROBE & 0.0319 & \bestresult{0.3536} & 0.0478 & 0.5652 & 6.20 & 0.3464 & -35.12 & 5.00 \\
\addlinespace[2pt]
MCTS & 0.0281 & 0.3238 & 0.0409 & 0.5145 & 5.68 & 0.3296 & -31.56 & 7.14 \\
Graph (Ours) & \bestresult{0.0324} & 0.3441 & \bestresult{0.0494} & \bestresult{0.5670} & \bestresult{8.44} & \bestresult{0.5332} & \secondresult{-25.07} & \bestresult{1.43} \\
\bottomrule
\end{tabular*}
\end{table}

\begin{table}[t]
\caption{CSI 500 test performance under the protocol and notation of Table~\ref{tab:main-csi300}. Higher is better except for mean rank; signed MDD closer to zero is better.}
\label{tab:main-csi500}
\centering
\small
\setlength{\tabcolsep}{4pt}
\renewcommand{\arraystretch}{1.15}
\begin{tabular*}{\linewidth}{@{\extracolsep{\fill}}lrrrrrrrr@{}}
\toprule
\multirow{2}{*}{Method} & \multicolumn{4}{c}{Predictive} & \multicolumn{3}{c}{Portfolio} & \multirow{2}{*}{Mean rank} \\
\cmidrule(lr){2-5} \cmidrule(lr){6-8}
& IC & ICIR & RankIC & RankICIR & ARR & IR & MDD & \\
\midrule
Alpha158 & 0.0339 & \secondresult{0.3927} & \secondresult{0.0618} & \bestresult{0.7898} & 11.07 & 0.5856 & -28.41 & 4.43 \\
GP & 0.0213 & 0.2752 & 0.0438 & 0.6092 & 12.52 & 0.6101 & -38.06 & 8.57 \\
AlphaGen & 0.0296 & 0.3613 & 0.0573 & 0.7147 & 11.82 & 0.6077 & -34.26 & 6.43 \\
AlphaForge & 0.0284 & 0.3129 & 0.0590 & 0.7076 & 11.08 & 0.5758 & -28.70 & 6.64 \\
AlphaSAGE & 0.0245 & 0.2652 & 0.0531 & 0.6394 & 10.51 & 0.5533 & -27.69 & 8.71 \\
AlphaAgent & 0.0227 & 0.2516 & 0.0522 & 0.5917 & \secondresult{12.76} & \bestresult{0.6798} & \secondresult{-26.51} & 6.71 \\
R\&D-Agent & 0.0236 & 0.2679 & 0.0543 & 0.6855 & 11.88 & 0.6239 & -33.44 & 7.43 \\
QuantaAlpha & 0.0266 & 0.3221 & 0.0533 & 0.6629 & 12.16 & 0.6370 & -29.20 & 6.07 \\
AlphaPROBE & 0.0309 & 0.3688 & 0.0585 & 0.7551 & 11.37 & 0.5937 & -29.98 & 5.57 \\
\addlinespace[2pt]
MCTS & \secondresult{0.0356} & 0.3898 & 0.0590 & 0.7472 & 12.16 & 0.6258 & -32.18 & \secondresult{4.14} \\
Graph (Ours) & \bestresult{0.0360} & \bestresult{0.3997} & \bestresult{0.0627} & \secondresult{0.7562} & \bestresult{13.01} & \secondresult{0.6689} & \bestresult{-26.42} & \bestresult{1.29} \\
\bottomrule
\end{tabular*}
\end{table}

\subsection{Ablation Study}
\label{sec:ablation}

We use the CSI 300 ablations to isolate five parts of the search: Graph Flow
propagation, intrinsic-quality archive construction, parent ranking after Flow,
utility composition, and graph relations. The propagation control sets
$\gamma=0$. Quality scalarization replaces the intrinsic-quality Pareto archive
with an equal-weight scalar score; this replacement also changes the
high-quality reference set and the Fertility computation used by the search.
Flow scalarization keeps Graph Flow but replaces Pareto fronts and crowding
distance for parent ranking with an equal-weight score over normalized
propagated $(Q,F,D)$. The utility controls remove Fertility, remove Diversity,
or retain Quality alone, and the relation controls remove Generation,
Structure, or Correlation individually. The latter two remove the relation
from both Flow and AverageDistance; Appendix~\ref{app:relation-role-controls}
separates these roles. Here $Q$, $F$, and $D$ denote
SelfQuality, Fertility, and AverageDistance, while $G$, $S$, and $C$ denote the
three relations. All variants use the same data split, LLM and prompt
configuration, search budget, and downstream evaluation protocol.

Every ablation in Table~\ref{tab:ablation} falls below the full model on all
seven metrics. This consistent pattern supports the joint use of the quality
archive, Flow-level parent selection, all three utility objectives, and all
three graph relations. Appendix~\ref{app:ablation-semantics} gives the detailed
changes to the quality reference set, Fertility, Flow normalization, and
relation handling.

\begin{table}[t]
\caption{Ablation study on CSI 300 after LightGBM integration. Quality scalarization changes the intrinsic-quality archive, reference-set construction, and Fertility computation; Flow scalarization changes parent ranking after Graph Flow. Remaining variants remove the named utility or relation. ARR and signed MDD are percentages; higher is better for all seven metrics. Bold marks the best value in each column.}
\label{tab:ablation}
\centering
\small
\setlength{\tabcolsep}{3pt}
\renewcommand{\arraystretch}{1.15}
\begin{tabular*}{\linewidth}{@{\extracolsep{\fill}}>{\raggedright\arraybackslash}p{0.40\linewidth}rrrrrrr@{}}
\toprule
\multirow{2}{*}{Variant} & \multicolumn{4}{c}{Predictive} & \multicolumn{3}{c}{Portfolio} \\
\cmidrule(lr){2-5} \cmidrule(lr){6-8}
& IC & ICIR & RankIC & RankICIR & ARR & IR & MDD \\
\midrule
Full Graph: $(Q,F,D)$; relations $(G,S,C)$ & \bestresult{0.0324} & \bestresult{0.3441} & \bestresult{0.0494} & \bestresult{0.5670} & \bestresult{8.44} & \bestresult{0.5332} & \bestresult{-25.07} \\
\textit{w/o} Flow: $\gamma=0$ & 0.0278 & 0.2929 & 0.0436 & 0.4937 & 5.90 & 0.3541 & -28.73 \\
Quality scalarization: intrinsic quality & 0.0284 & 0.3107 & 0.0452 & 0.5117 & 5.80 & 0.3539 & -27.13 \\
Flow scalarization: normalized $(Q,F,D)$ & 0.0289 & 0.2987 & 0.0474 & 0.5201 & 5.80 & 0.3486 & -26.86 \\
\midrule
\textit{w/o} Fertility: Pareto $(Q,D)$ & 0.0297 & 0.3181 & 0.0456 & 0.5286 & 6.71 & 0.4132 & -26.35 \\
\textit{w/o} Diversity: Pareto $(Q,F)$ & 0.0286 & 0.3119 & 0.0448 & 0.5252 & 6.95 & 0.4273 & -26.16 \\
Quality only: Pareto $(Q)$ & 0.0296 & 0.3099 & 0.0475 & 0.5340 & 5.64 & 0.3421 & -27.33 \\
\midrule
\textit{w/o} Generation relation & 0.0298 & 0.3189 & 0.0470 & 0.5499 & 6.59 & 0.3907 & -28.55 \\
\textit{w/o} Structure relation & 0.0276 & 0.3022 & 0.0453 & 0.5289 & 4.37 & 0.2659 & -29.27 \\
\textit{w/o} Correlation relation & 0.0295 & 0.3290 & 0.0447 & 0.5304 & 7.28 & 0.4401 & -26.06 \\
\bottomrule
\end{tabular*}
\end{table}

\subsection{Search Behavior Analysis}
\label{sec:search-behavior}

\begin{figure}[!htbp]
\centering
\includegraphics[width=\textwidth]{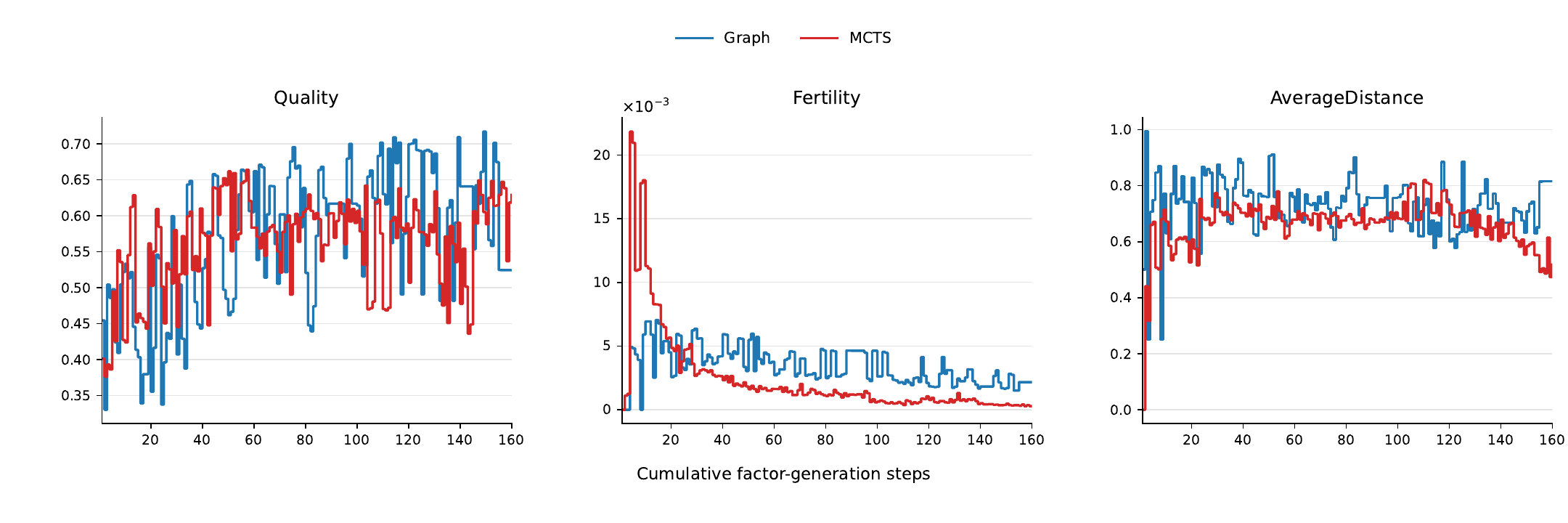}
\caption{Search-state trajectories of CosmosAlpha-Graph and
CosmosAlpha-MCTS over cumulative factor-generation steps. From left to right,
the panels show Flow-propagated Quality, Fertility, and AverageDistance on
native vertical scales. MCTS coordinates are offline diagnostics.}
\label{fig:search-behavior}
\end{figure}

Figure~\ref{fig:search-behavior} compares three complementary search diagnostics:
Quality combines predictive strength, temporal stability, and aligned return;
Fertility reflects historical frontier gains from parent expansions;
AverageDistance measures structural and correlation novelty relative to the
high-quality reference set. All panels show Flow-propagated
coordinates of newly generated factors after their generation batch. A newborn
has zero raw Fertility but may receive positive propagated Fertility from its
neighbors. For MCTS, we replay logged expansions, reconstruct the three relations,
and apply the same diagnostics and Flow settings ($\gamma=0.75$, equal relation
weights) offline. In this replay, birth-time hypervolume gains are computed
against the frozen pre-event archive. These diagnostics do not enter MCTS
selection or backup.

MCTS often has higher Quality in the first 20 steps. Both Quality trajectories
fluctuate thereafter, with Graph often reaching higher values in the middle
and later stages. MCTS's propagated Fertility declines toward zero after its
early peak, whereas Graph retains a comparatively steady, positive signal
later in search. This relatively stable Fertility is consistent with Graph's
overall improvement in Quality. Graph also has higher propagated
AverageDistance over most of the
trajectory, consistent with greater structural and correlation novelty
relative to its high-quality reference set.

\subsection{Additional Analyses}
Under shared generation conditions and a 4M-Token cap, Graph leads MCTS and Chain
on all seven metrics (Appendix~\ref{app:matched-budget}). Wider normalization
intervals retain comparable performance, and relation-role controls support
both propagation and diversity (Appendices~\ref{app:normalization-sensitivity}
and~\ref{app:relation-role-controls}). Five rolling experiments repeat factor
mining and model fitting; Graph exceeds Alpha158 in RankIC and ARR in four
test years each (Appendix~\ref{app:rolling-robustness}). Transaction-cost tests
retain Graph's relative advantage, including over the CSI 300 index at the
highest tested cost (Appendix~\ref{app:transaction-costs}).

\section{Conclusion}
\label{sec:conclusion}

We introduced CosmosAlpha, an LLM-driven framework that organizes evaluated
factors as a multi-relational search state. It connects generation lineage,
expression structure, and cross-sectional exposure correlation, propagates
quality, fertility, and diversity through Graph Flow, and uses Pareto-guided
selection to guide expansion beyond local generation history.

Under a common LightGBM integration protocol, CosmosAlpha-Graph achieves the
best mean rank on both CSI 300 and CSI 500 and outperforms the controlled
CosmosAlpha-MCTS variant on every reported metric. Every tested ablation
underperforms the full model across all seven metrics, supporting relational
propagation, Pareto-guided selection, utility composition, and the three graph
relations. The LLM and independent-run studies further show that this advantage
is not confined to one backbone or stochastic run. Future work will apply
CosmosAlpha to additional markets and to the discovery of high-frequency and
fundamental factors.

\subsection*{AI use statement}

Generative AI tools were used primarily for English-language editing, grammar
correction, and local improvements in clarity and organization. The research
question, core ideas, method design, experimental protocol, implementation,
analysis, and conclusions were developed by the authors. The authors reviewed
and verified all AI-assisted text, equations, citations, technical
descriptions, and reported results, and take full responsibility for the final
manuscript.

\subsection*{Ethics statement}

This work uses historical market data and does not involve human participants,
personal data, or interaction with individual investors. The study is an
offline research evaluation, and its backtest results should not be interpreted
as financial advice or guaranteed live-trading performance. Automated factor
discovery can encourage overfitting or unsafe deployment when used without
independent validation and appropriate risk controls. We therefore report the
temporal split and held-out
test protocol and limit our claims to the evaluated stock universes and period.

\subsection*{Reproducibility statement}

Appendix~\ref{app:experiment-details} specifies the rolling-operator lookback
window, temporal splits, factor-pool construction, preprocessing, LightGBM
configuration, and portfolio protocol. It also states that factor mining and
LightGBM early stopping use only the training and validation periods,
respectively, while all reported metrics are computed on the held-out test
period. Appendix~\ref{app:llm-sensitivity} and
Appendix~\ref{app:run-stability} document the tested LLM identifiers and
independent runs.
The released artifacts will identify the exact code revision used for each
experiment and provide the environment specification, prompts, model
identifiers, run configurations, and scripts needed to regenerate the reported
tables and figures.

\section*{Acknowledgments}

This work is supported by the National Natural Science Foundation of China
(Grant No.~12475064), the National Key Research and Development Program of China
(Grant No.~2021YFC2203004), and the Fundamental Research Funds for the Central
Universities.

\bibliography{references}
\bibliographystyle{plainnat}

\clearpage
\appendix

\section{Additional Method Details}
\label{app:method-details}

This appendix specifies the search semantics omitted from
Section~\ref{sec:method}. We focus on choices that change graph state, node
utility, or the search trajectory, and report mechanism-defining numerical
settings where they clarify the algorithm. Metric definitions, prompt
templates, and the factor-language catalog are kept separate from these
algorithmic definitions. Appendix~\ref{app:metrics} defines the single-factor
statistics used by the search.

\subsection{Initialization and Node Admission}
\label{app:initialization-admission}

Each run begins with an initial set of factors proposed by the LLM. These
proposals pass through the same expression validation, factor calculation, and
evaluation pipeline as factors generated in later rounds. Once the initial
factors have been admitted, CosmosAlpha constructs their relation matrices and
initializes the Quality Pareto Archive before selecting any parent for
expansion.

Graph admission is a technical-validity condition rather than a
quality-selection rule. A candidate must have a valid expression, successfully
computed factor values, a completed evaluation under the run's common data and
preprocessing protocols, and finite values for mean RankIC, RankIC standard
deviation, and direction-aligned Top-20\% return. The evaluation must also
contain at least two valid dates and cover at least 80\% of the candidate
evaluation dates. Candidates failing these conditions do not become graph nodes;
quality comparisons are applied only after admission through the Quality
Pareto Archive and parent-selection procedure.

\subsection{Canonical Factor Identity and Duplicate Handling}
\label{app:factor-identity}

CosmosAlpha assigns graph identity from a canonicalized expression under a
fixed data frequency and factor language. Canonicalization recursively
normalizes the abstract syntax tree (AST) and orders the operands of supported
commutative operators listed in Table~\ref{tab:operator-catalog}, while
retaining operator identity, tree topology, fields, constants, and window
arguments. It therefore removes equivalent operand orderings without
attempting general algebraic simplification.

Duplicate checking occurs after expression validation and before factor
calculation. Each validated candidate is compared with all committed graph
nodes and with the other candidates proposed in the same batch. If several
same-batch candidates have the same canonical identity, one is retained by a
deterministic ordering and the rest are treated as duplicates. Expressions
that have already been evaluated but failed the graph-admission conditions are
likewise not reevaluated. A duplicate candidate creates neither a node nor a
generation edge, but its parent expansion still consumes one attempt; an
attempt with no valid unique child contributes zero to Fertility.

\subsection{Multi-Relational Graph Construction}
\label{app:graph-construction}

The three relation matrices are reconstructed over the currently admitted
nodes. Generation records search provenance, whereas structure and correlation
first define pairwise similarities and then project them into local
neighborhoods for Graph Flow.

\paragraph{Generation relation.}
Every admitted child records the single selected factor from which it was
mutated. The stored provenance is directed, but Graph Flow uses a symmetric
unit-weight edge between each direct parent--child pair. No explicit edges are
added between more distant ancestors and descendants; their influence arises
through repeated propagation. Invalid or duplicate proposals do not create
generation edges.

\paragraph{Structure relation.}
For each factor, CosmosAlpha extracts the set of canonical subexpressions from
its AST. A subexpression preserves its operator, child topology, primitive
fields, constants, and window arguments, and commutative operands are ordered
before extraction. Structural similarity is the Jaccard similarity in
Eq.~\ref{eq:structure-similarity}; because the representation is a set,
repeated occurrences of an identical subtree are counted once.

\paragraph{Correlation relation.}
Correlation is calculated from the evaluated and preprocessed stock exposures,
not from raw expression outputs. For a factor pair, the calculation first
requires the two exposure histories to overlap on at least 80\% of the
evaluation calendar. On each overlapping date, Pearson correlation is computed
over stocks for which both exposures are finite and eligible. A daily value is
retained only if this intersection contains at least 100 stocks and covers at
least 80\% of that day's eligible universe. The remaining daily correlations
are averaged with equal date weights as in Eq.~\ref{eq:factor-corr}. A pair can
form an edge only when valid daily correlations cover at least 80\% of the full
evaluation calendar and at least 80\% of its overlapping dates. The signed value
$C_{ij}$ is retained as a correlation fact, while relation strength and
distance use $|C_{ij}|$ and $1-|C_{ij}|$, respectively.

\paragraph{Local neighborhood projection.}
Structure candidates must satisfy $S^{\mathrm{str}}_{ij}\geq 0.20$, whereas
reliable correlation candidates must satisfy $|C_{ij}|\geq 0.10$. For each
relation separately, a node ranks the remaining candidates by similarity and
retains at most its Top-30 neighbors. The final undirected graph uses union
symmetrization: edge $(i,j)$ is retained if either $i$ selects $j$ or $j$
selects $i$, with weight $S^{\mathrm{str}}_{ij}$ or $|C_{ij}|$. A node's final
undirected degree can therefore exceed 30. Before the three modalities are
combined, each is normalized independently; a row with no retained neighbor is
replaced by a self-loop.

\subsection{Archive-Based Search Utilities}

\paragraph{Quality normalization and archive.}
\label{app:quality-details}
The Quality Pareto Archive applies dominance to the unclipped raw vector in
Eq.~\ref{eq:quality-vector}. Fixed anchors are used only to construct scalar
node Quality and to calculate hypervolume. The normalization ranges are
$[0,0.05]$ for absolute mean RankIC, $[0.05,0.09]$ in the reverse direction for
RankIC standard deviation, and $[0,8\times10^{-4}]$ for direction-aligned
Top-20\% return. The resulting coordinates are
\begin{align}
\widetilde q_i^{\mathrm{IC}}
&=\operatorname{clip}\!\left(
\frac{|\overline{\mathrm{RankIC}}_i|}{0.05},0,1\right),\\
\widetilde q_i^{\mathrm{std}}
&=\operatorname{clip}\!\left(
\frac{0.09-\sigma(\mathrm{RankIC}_i)}{0.09-0.05},0,1\right),\\
\widetilde q_i^{\mathrm{ret}}
&=\operatorname{clip}\!\left(
\frac{R_i^{\mathrm{Top20}}}{8\times10^{-4}},0,1\right).
\end{align}
Node Quality is their equal-weight mean,
\begin{equation}
Q_i^{\mathrm{self}}=
\frac{\widetilde q_i^{\mathrm{IC}}+
\widetilde q_i^{\mathrm{std}}+
\widetilde q_i^{\mathrm{ret}}}{3}.
\end{equation}
The anchors are fixed before search and are never estimated from a current
batch or graph. Hypervolume is calculated from the same normalized coordinates
in $[0,1]^3$ with reference point $(0,0,0)$; archive dominance itself remains
defined on the raw objectives.

The clipping bounds are empirical reference levels for strong single-factor
performance. Clipping limits rewards beyond these levels to encourage balanced
improvement. Appendix~\ref{app:normalization-sensitivity} reports clipping
frequencies and sensitivity to alternative normalizations.

\paragraph{Fertility.}
\label{app:fertility-details}
At the start of round $r$, the current Quality Pareto Archive is fixed as
$\mathcal A_r^-$. Every valid unique child in that round receives the
birth-time gain in Eq.~\ref{eq:delta-hv} relative to this same archive. These
gains are computed independently, so neither completion order nor a sibling's
quality changes another child's reward. The archive is updated jointly only
after all birth-time gains have been assigned.

An expansion attempt may produce more than one valid child. If
$\mathcal C_e$ is the set of valid unique children from attempt $e$, its
productive outcome is
\begin{equation}
y_e=\begin{cases}
\max_{c\in\mathcal C_e}\Delta HV_c,&\mathcal C_e\neq\emptyset,\\
0,&\mathcal C_e=\emptyset.
\end{cases}
\end{equation}
Thus, failed, invalid, duplicate-only, and otherwise childless attempts all
contribute zero. Fertility operationalizes a parent's productive potential as
the mean of its three best attempt outcomes,
\begin{equation}
Q_i^{\mathrm{fert}}=\operatorname{Top3Mean}
(y_{i,1},y_{i,2},\ldots),
\end{equation}
where missing outcomes are padded with zero when fewer than three attempts have
been observed.

\paragraph{Diversity reference set.}
\label{app:diversity-reference}
At the beginning of each round, a high-quality reference set $\mathcal R_r$ is
selected from the current Quality Pareto Archive. This set is held fixed
throughout the round: children generated in round $r$ are compared with
$\mathcal R_r$ and can enter the reference set only from the following round
onward. We use $K=10$ for both relation spaces. For each relation
$m\in\{\mathrm{str},\mathrm{corr}\}$, the node itself is excluded and the
closest available reference factors define
\begin{equation}
D_i^m=\frac{1}{|\mathcal N_K^m(i)|}
\sum_{j\in\mathcal N_K^m(i)}d_{ij}^m,
\qquad
Q_i^{\mathrm{div}}=\frac{D_i^{\mathrm{str}}+D_i^{\mathrm{corr}}}{2}.
\end{equation}
If fewer than $K$ references are available for a modality, all available
references are used; if none remains after self-exclusion and validity checks,
that modality contributes zero. Generation distance is excluded because
lineage is already represented directly by the generation graph.

\subsection{Graph Flow and Pareto Parent Selection}

The relation matrices are row-normalized independently and combined as in
Eq.~\ref{eq:multi-relation-transition}. The self-loop assigned to an empty
modality row preserves that node's own signal for that modality. In particular,
if a node is isolated under all three relations, its combined transition row is
a self-loop and Graph Flow leaves its utility equal to its raw utility. Starting
from $F^{(0)}=R$, the implementation evaluates Eq.~\ref{eq:graph-flow} with
$\gamma=0.75$. Fixed-point iteration stops when the maximum absolute change in
any utility coordinate is at most $10^{-8}$ or after 100 iterations.

A node is eligible for expansion while it has made fewer than five previous
attempts. Eligible nodes are sorted into non-dominated fronts in the propagated
utility space
$\mathbf F_i=(F_i^{\mathrm{self}},F_i^{\mathrm{fert}},F_i^{\mathrm{div}})$.
Up to four parents are taken without replacement from successive fronts in each
round. If the next front exceeds the remaining slots, standard NSGA-II crowding
distance prioritizes less crowded points~\citep{deb2002nsga2}; any remaining
ties are resolved deterministically. Diversity therefore enters parent selection
through the propagated AverageDistance objective and, when a front must be
truncated, through crowding in the three-dimensional utility space. The selected
batch is the minimum of four, the number of eligible nodes, and the remaining
global budget, so the final batch may be smaller than earlier batches.

The global budget counts parent expansion attempts. Each selected parent makes
one attempt in the round, and success, duplication, expression invalidity,
evaluation failure, and LLM failure all consume one unit. Expression-repair
calls within that attempt do not consume additional graph-search budget.

\subsection{Parent-Grounded Mutation}
\label{app:parent-grounded-mutation}

Each expansion is conditioned on one selected parent rather than on an
undifferentiated collection of prior factors. The mutation context combines the
parent expression with its raw quality metrics, current Quality, Fertility,
AverageDistance, expansion history, objective-wise gaps to the current archive,
and correlation-neighborhood constraints. This context directs the LLM to
propose up to two structural or signal-level mutations of the parent instead of
unrelated factors.

The resulting designs are converted into expressions under the shared factor
language and then pass through validation, optional repair, batch-level
duplicate checking, factor calculation, and evaluation. Only valid unique
candidates satisfying the node-admission conditions are inserted into the
graph. The current implementation uses single-parent mutation; it does not
combine two parents through crossover.

\subsection{Round-Level Search Procedure}
\label{app:search-details}

Algorithm~\ref{alg:cosmosalpha} makes the round boundaries explicit. The key
invariant is that both the Quality Pareto Archive and the high-quality reference
set are fixed before parent selection and remain unchanged while all attempts
in the batch are evaluated.

Within a batch, deterministic duplicate resolution prevents candidate
completion order from deciding which equivalent proposal is retained.
Likewise, evaluating all children against $\mathcal A_r^-$ makes birth-time
hypervolume gains independent of child completion order. At the next round
boundary, the updated archive determines a new reference set and all node
utilities are refreshed before parent selection.

\begin{algorithm}[t]
\caption{CosmosAlpha Multi-Relational Graph Search}
\label{alg:cosmosalpha}
\begin{algorithmic}[1]
\REQUIRE LLM, common factor evaluator, expansion-attempt budget $B$
\ENSURE Evaluated factor graph $\mathcal G$ and Quality Pareto Archive $\mathcal A$
\STATE Generate, validate, deduplicate, and evaluate the initial factors
\STATE Admit technically valid factors; construct $\mathcal G$ and $\mathcal A$
\WHILE{fewer than $B$ expansion attempts have been used}
    \STATE Fix the pre-round archive $\mathcal A_r^-$ and reference set $\mathcal R_r$
    \STATE Compute AverageDistance to $\mathcal R_r$ for all current nodes
    \STATE Build $P$; run Graph Flow from the current vector utilities
    \STATE Select an eligible parent batch by Pareto fronts and crowding distance
    \IF{no eligible parent can be selected}
        \STATE \textbf{break}
    \ENDIF
    \FOR{each selected parent}
        \STATE Generate parent-grounded mutations with the LLM
        \STATE Validate candidates and resolve committed and same-batch duplicates
        \STATE Calculate and evaluate the surviving candidates
    \ENDFOR
    \STATE Compute every valid child's $\Delta HV$ against $\mathcal A_r^-$
    \STATE Record attempt outcomes and update parent Fertility
    \STATE Commit all valid unique children jointly to $\mathcal A$
    \STATE Insert the children and update generation, structure, and correlation relations
    \STATE Compute new-node diversity against the fixed set $\mathcal R_r$
\ENDWHILE
\RETURN $\mathcal G,\mathcal A$
\end{algorithmic}
\end{algorithm}

\section{Metrics and Evaluation Definitions}
\label{app:metrics}

This appendix defines the quantities used for search-time factor evaluation and
for the final out-of-sample comparison. Following the experimental protocol in
Appendix~\ref{app:experiment-details}, search-time statistics are computed only
on the training period, whereas integrated predictive and portfolio metrics are
computed on the held-out test period. All temporal averages below assign equal
weight to each valid trading date.

\subsection{Single-Factor Metrics}
\label{app:single-factor-metrics}

For factor $f_i$ on date $t$, let $z_{i,j,t}$ denote its processed exposure for
stock $j$, and let $y_{j,t}$ be the corresponding raw one-day forward return.
The valid cross-section $\mathcal U_{i,t}$ contains eligible stocks for which
both quantities are finite, with $n_{i,t}=|\mathcal U_{i,t}|$. A date is retained
when $n_{i,t}\geq100$ and the valid observations cover at least 80\% of the
eligible universe on that date. In addition, a factor admitted as a graph node
must have at least two valid dates covering at least 80\% of the candidate
evaluation dates.

For two cross-sectional vectors $a$ and $b$ over a valid set $\mathcal U$, we
write their Pearson correlation as
\begin{equation}
\operatorname{Corr}_{\mathcal U}(a,b)=
\frac{\sum_{j\in\mathcal U}(a_j-\bar a)(b_j-\bar b)}
{\sqrt{\sum_{j\in\mathcal U}(a_j-\bar a)^2}
 \sqrt{\sum_{j\in\mathcal U}(b_j-\bar b)^2}}.
\label{eq:appendix-cross-sectional-corr}
\end{equation}
The daily information coefficient and rank information coefficient are
\begin{align}
\mathrm{IC}_{i,t}
&=\operatorname{Corr}_{\mathcal U_{i,t}}(z_{i,\cdot,t},y_{\cdot,t}),
\label{eq:appendix-daily-ic}\\
\mathrm{RankIC}_{i,t}
&=\operatorname{Corr}_{\mathcal U_{i,t}}
\bigl(\operatorname{rank}(z_{i,\cdot,t}),
      \operatorname{rank}(y_{\cdot,t})\bigr),
\label{eq:appendix-daily-rankic}
\end{align}
where tied observations receive their average rank. Dates with a degenerate
cross-section, for which either correlation is undefined, are excluded.

Let $\mathcal T_i$ be the factor's valid-date set and $T_i=|\mathcal T_i|$.
For $x_{i,t}\in\{\mathrm{IC}_{i,t},\mathrm{RankIC}_{i,t}\}$, its time-series
mean, sample standard deviation, and information ratio are
\begin{align}
\bar x_i&=\frac{1}{T_i}\sum_{t\in\mathcal T_i}x_{i,t},\\
s(x_i)&=\sqrt{\frac{1}{T_i-1}
\sum_{t\in\mathcal T_i}(x_{i,t}-\bar x_i)^2},\\
\mathrm{IR}(x_i)&=\frac{\bar x_i}{s(x_i)}.
\label{eq:appendix-factor-aggregation}
\end{align}
The temporal means retain their signs; the absolute value is introduced only
when mean RankIC enters the Quality vector in Eq.~\ref{eq:quality-vector}.
Accordingly, ICIR and RankICIR are the unannualized ratios obtained from the
daily IC and RankIC series. The search additionally uses
$s(\mathrm{RankIC}_i)$ as its RankIC standard deviation. If fewer than two
valid dates are available, or if the sample standard deviation is
numerically zero, the corresponding ratio is set to zero; the graph-admission
rule separately excludes factors without two valid dates.

The Top-20\% return aligns each factor before forming the upper-exposure group.
Its orientation is fixed from the factor's mean RankIC,
\begin{equation}
o_i=\begin{cases}
+1,&\overline{\mathrm{RankIC}}_i\geq0,\\
-1,&\overline{\mathrm{RankIC}}_i<0.
\end{cases}
\label{eq:appendix-factor-orientation}
\end{equation}
Let $k_{i,t}=\lfloor0.2n_{i,t}\rfloor$ and let $\mathcal H_{i,t}$ contain the
$k_{i,t}$ stocks with the largest values of $o_i z_{i,j,t}$. Exposure ties at
the selection boundary are resolved by instrument identifier. The daily and
time-averaged returns are
\begin{equation}
r^{\mathrm{Top20}}_{i,t}
=\frac{1}{k_{i,t}}\sum_{j\in\mathcal H_{i,t}}y_{j,t},
\qquad
R_i^{\mathrm{Top20}}
=\frac{1}{T_i}\sum_{t\in\mathcal T_i}r^{\mathrm{Top20}}_{i,t}.
\label{eq:appendix-top20-return}
\end{equation}
This quantity is a search-time single-factor statistic based on raw forward
returns; it is distinct from the downstream TopK-Drop portfolio backtest.

\subsection{Integrated Predictive Metrics}
\label{app:integrated-predictive-metrics}

The final predictive metrics use the same daily definitions but evaluate the
integrated LightGBM signal rather than an individual factor. Specifically, the
factor exposure in Eqs.~\ref{eq:appendix-daily-ic}
and~\ref{eq:appendix-daily-rankic} is replaced by the model prediction
$\hat y_{j,t}$, while the correlation target remains the raw one-day forward
return $y_{j,t}$. A test date is valid when at least 100 stocks have finite
predictions and returns. The reported IC and RankIC are the signed means of the
valid daily correlations; ICIR and RankICIR divide those means by their sample
standard deviations as in Eq.~\ref{eq:appendix-factor-aggregation}, without
annualization or taking absolute values.

\subsection{Portfolio Metrics}
\label{app:portfolio-metrics}

The LightGBM signal is converted into daily holdings by the Qlib TopK-Drop
backtest described in Appendix~\ref{app:experiment-details}. Let $r_t$ be its
daily absolute portfolio return; neither a market benchmark nor a risk-free
rate is subtracted.
Consistent with Qlib's daily arithmetic-return analysis~\citep{yang2020qlib},
let $A=238$, and define
\begin{equation}
\bar r=\frac{1}{T}\sum_{t=1}^{T}r_t,
\qquad
s_r=\sqrt{\frac{1}{T-1}\sum_{t=1}^{T}(r_t-\bar r)^2}.
\end{equation}
Annualized return and portfolio information ratio are then
\begin{equation}
\mathrm{ARR}=A\bar r,
\qquad
\mathrm{IR}=\sqrt{A}\,\frac{\bar r}{s_r}.
\label{eq:appendix-portfolio-arr-ir}
\end{equation}
For signed maximum drawdown, define the arithmetic cumulative return
$C_t=\sum_{u=1}^{t}r_u$. We report
\begin{equation}
\mathrm{MDD}=\min_{1\leq t\leq T}
\left(C_t-\max_{1\leq u\leq t}C_u\right)\leq0.
\label{eq:appendix-signed-mdd}
\end{equation}
Thus, a larger signed MDD (closer to zero) indicates a smaller drawdown. ARR and
MDD are multiplied by 100 when displayed as percentages in the tables.

For method $m$, the final Mean Rank summarizes the four predictive and three
portfolio metrics:
\begin{equation}
\mathrm{MeanRank}_m=\frac{1}{7}\sum_{q=1}^{7}\operatorname{rank}_{m,q}.
\label{eq:appendix-mean-rank}
\end{equation}
Rank 1 is best and ties receive their average rank. All seven columns are ranked
in descending order because MDD is signed; hence a smaller Mean Rank is better.

\subsection{Repeated-Run Aggregation}
\label{app:repeated-run-metrics}

The stability analysis uses $L=3$ independent LLM runs under the same data,
model, and search configuration. For a reported metric value $v_\ell$ from run
$\ell$, the Mean and Std. rows are
\begin{equation}
\bar v=\frac{1}{L}\sum_{\ell=1}^{L}v_\ell,
\qquad
\sigma_{\mathrm{run}}=
\sqrt{\frac{1}{L}\sum_{\ell=1}^{L}(v_\ell-\bar v)^2}.
\label{eq:appendix-run-aggregation}
\end{equation}
The reported run-level dispersion therefore uses the population standard
deviation across the three observed runs. It should not be confused with the
sample standard deviation across trading dates used by ICIR, RankICIR, and
portfolio IR.

\section{Experimental Configuration}
\label{app:experiment-details}

This appendix consolidates the principal data, search, model, and backtest
settings used in the main experiments. Unless a row is explicitly marked as
Graph-specific or ablation-only, the CSI 300 and CSI 500 experiments use the
same setting. Metric formulas are defined separately in
Appendix~\ref{app:metrics}.

\subsection{Data, Temporal Splits, and Common Evaluation}

All methods are evaluated on the same daily A-share snapshot. Historical index
membership and trading-eligibility masks are applied point-in-time. Factor
mining and all search feedback use only the training period; the mined factor
pool is then fixed before downstream validation and testing. Table~\ref{tab:data-config}
summarizes the common data and factor-evaluation protocol.

\begin{table}[!htbp]
\caption{Shared data, preprocessing, and factor-evaluation configuration.}
\label{tab:data-config}
\centering
\footnotesize
\setlength{\tabcolsep}{4pt}
\renewcommand{\arraystretch}{1.12}
\begin{tabular}{>{\raggedright\arraybackslash}p{0.28\linewidth}>{\raggedright\arraybackslash}p{0.38\linewidth}>{\raggedright\arraybackslash}p{0.24\linewidth}}
\toprule
Setting & Value & Scope or provenance \\
\midrule
Universes & CSI 300 and CSI 500 with historical membership & Shared across methods \\
Frequency and prices & Daily; post-adjusted prices & Shared data protocol \\
Rolling lookback start & January 4, 2010 & Supports lagged operators \\
Mining and training & April 1, 2010--May 31, 2019 & Search feedback and LightGBM fitting \\
Validation & June 1, 2019--December 31, 2020 & LightGBM early stopping only \\
Test & January 1, 2021--December 31, 2025 & Final prediction and portfolio evaluation only \\
Return label & One-trading-day forward close-to-close return & Shared across methods \\
Stock eligibility & At least one year since listing; blacklist and tradability masks enabled; financial stocks retained & Shared data protocol \\
Rolling-data requirement & At least two thirds of a window; sample standard deviation uses $\mathrm{ddof}=1$ & Factor calculation \\
Cross-sectional processing & MAD winsorization with multiplier $5$, neutralization, then Z-score normalization & Shared across methods \\
Neutralization exposures & Industry, log market capitalization, 20-day return, turnover, and volatility & Shared across methods \\
Minimum daily sample & $100$ finite eligible stocks and at least $80\%$ daily coverage & Factor evaluation \\
Minimum date coverage & At least two valid dates covering $80\%$ of candidate evaluation dates & Graph node admission \\
\bottomrule
\end{tabular}
\end{table}

The downstream target is the same one-day return, cross-sectionally rank
normalized for LightGBM training. Predictions are evaluated against the raw
forward return as specified in Appendix~\ref{app:integrated-predictive-metrics}.
No additional pairwise-correlation filter is applied before model fitting.

\subsection{Search Budget and Expression Generation}

Main CosmosAlpha runs begin with factors generated within each run. Automated
methods target approximately 150 evaluable factors, balancing mining cost and
model input size near Alpha158. This output target does not equalize proposals,
repairs, evaluations, or Tokens. Appendix~\ref{app:matched-budget} separately
compares common initial factors and generation conditions under a 4M-Token cap.

\begin{table}[!htbp]
\caption{Search and expression-generation settings for CosmosAlpha-Graph.}
\label{tab:search-config}
\centering
\footnotesize
\setlength{\tabcolsep}{4pt}
\renewcommand{\arraystretch}{1.12}
\begin{tabular}{>{\raggedright\arraybackslash}p{0.34\linewidth}>{\raggedright\arraybackslash}p{0.20\linewidth}>{\raggedright\arraybackslash}p{0.36\linewidth}}
\toprule
Setting & Value & Role \\
\midrule
Initial Graph factors & $2$ & LLM-generated factors evaluated before the first expansion round \\
Parents per round & Up to $4$ & Parallel expansion width \\
Expansion cap per node & $5$ attempts & Prevents repeated use of one parent \\
Child proposals per attempt & Up to $2$ & Parent-grounded mutations \\
Factor-pool target & Approximately $150$ evaluable factors & Common output scale across automated methods \\
Maximum expression depth & $15$ & Expression-complexity constraint \\
Allowed rolling windows & $\{5,10,20\}$ trading days & Shared factor language \\
Expression-repair attempts & At most $2$ & Applied after an invalid generated expression \\
Duplicate policy & Reject an existing canonical expression & Shared factor-identity rule \\
\bottomrule
\end{tabular}
\end{table}

An expansion attempt remains part of the search budget when it yields an
invalid or duplicate proposal, as detailed in
Appendix~\ref{app:factor-identity}. Prompt text is provided in
Appendix~\ref{app:prompt-templates}; the complete factor language is specified
in Appendix~\ref{app:factor-language}.

\subsection{Graph Construction and Search Hyperparameters}

Tables~\ref{tab:graph-relation-config} and~\ref{tab:graph-utility-config}
collect the parameters that instantiate the Graph method described in
Section~\ref{sec:method}. Relation thresholds are applied before per-node
Top-$K$ truncation, and all values are fixed before a run begins.

\begin{table}[!htbp]
\caption{Graph relation and neighborhood settings.}
\label{tab:graph-relation-config}
\centering
\footnotesize
\setlength{\tabcolsep}{4pt}
\renewcommand{\arraystretch}{1.12}
\begin{tabular}{>{\raggedright\arraybackslash}p{0.33\linewidth}>{\raggedright\arraybackslash}p{0.20\linewidth}>{\raggedright\arraybackslash}p{0.37\linewidth}}
\toprule
Setting & Value & Role \\
\midrule
Structure edge threshold & $0.20$ & Minimum Jaccard similarity \\
Structure neighbors & Top-$30$ per node & Local structure neighborhood before union symmetrization \\
Correlation edge threshold & $|C_{ij}|\geq0.10$ & Minimum exposure-correlation strength \\
Correlation neighbors & Top-$30$ per node & Local correlation neighborhood before union symmetrization \\
Correlation daily validity & $100$ stocks and $80\%$ coverage & Reliable daily cross-sectional correlation \\
Correlation date validity & $80\%$ of the evaluation calendar and $80\%$ of overlapping dates & Reliable pairwise relation \\
AverageDistance neighbors & $K=10$ per relation & Nearest high-quality references in structure and correlation spaces \\
Generation edge weight & $1$ & Direct parent--child connection \\
\bottomrule
\end{tabular}
\end{table}

\begin{table}[!htbp]
\caption{Quality, Fertility, Graph Flow, and parent-selection settings.}
\label{tab:graph-utility-config}
\centering
\footnotesize
\setlength{\tabcolsep}{4pt}
\renewcommand{\arraystretch}{1.12}
\begin{tabular}{>{\raggedright\arraybackslash}p{0.34\linewidth}>{\raggedright\arraybackslash}p{0.22\linewidth}>{\raggedright\arraybackslash}p{0.34\linewidth}}
\toprule
Setting & Value & Role \\
\midrule
$|\overline{\mathrm{RankIC}}|$ anchors & $[0,0.05]$ & Quality and hypervolume normalization \\
RankIC standard-deviation anchors & $[0.05,0.09]$, reversed & Quality and hypervolume normalization \\
Top-20\% return anchors & $[0,8\times10^{-4}]$ & Quality and hypervolume normalization \\
Hypervolume reference point & $(0,0,0)$ & Quality Pareto Archive \\
Fertility history & Top-$3$ attempt outcomes & Productive-potential aggregation \\
Relation weights & $1/3$ each & Equal generation, structure, and correlation contribution \\
Graph Flow coefficient & $\gamma=0.75$ & Neighbor-propagation strength \\
Graph Flow solver & At most $100$ iterations; tolerance $10^{-8}$ & Fixed-point convergence \\
Parent selection & Pareto fronts with crowding distance & Multi-objective selection \\
Crowding tolerance & $10^{-12}$ & Degenerate-objective handling \\
\bottomrule
\end{tabular}
\end{table}

\subsection{LLM Configuration}

DeepSeek-V4-Flash is the default backbone for the main comparison. The prompt
renderer permits at most $32{,}000$ input tokens, retains up to four history
items, and requests at most $12{,}000$ output tokens. At most two formatting
retries are allowed. Table~\ref{tab:llm-config} lists the model-side settings;
the GPT and Kimi rows apply only to the sensitivity study in
Appendix~\ref{app:llm-sensitivity}.

\begin{table}[H]
\caption{LLM settings used in the main and backbone-sensitivity experiments.}
\label{tab:llm-config}
\centering
\footnotesize
\setlength{\tabcolsep}{5pt}
\begin{tabular}{llll}
\toprule
Use & Model identifier & Temperature & Reasoning control \\
\midrule
Main & DeepSeek-V4-Flash & $0.0$ & Maximum \\
Sensitivity & GPT-5.6-Sol & $0.0$ & xhigh \\
Sensitivity & Kimi K3 & $1.0$ & Maximum \\
\bottomrule
\end{tabular}
\end{table}

\paragraph{Observed LLM usage.}
With a budget of $100$ Graph expansion attempts, average per-run usage was
$6.17$ hours of cumulative LLM call time and $3.98$M tokens ($1.56$M input and
$2.41$M output) for DeepSeek-V4-Flash; $6.87$ hours and $1.42$M tokens ($0.96$M
input and $0.46$M output) for Kimi K3; and $2.18$ hours and $1.72$M tokens
($1.25$M input and $0.47$M output) for GPT-5.6-Sol.
Cumulative call time sums recorded client-call durations rather than end-to-end
wall-clock time. Token totals are lower bounds because some terminal records
lack provider-reported counts.

\subsection{LightGBM and Portfolio Configuration}

All factor pools are recomputed on the common data snapshot and passed to the
same LightGBM and Qlib backtest pipeline. The LightGBM settings follow the
Qlib Alpha158 benchmark~\citep{yang2020qlib}. Table~\ref{tab:downstream-config}
reports the principal settings that affect the displayed predictive and
portfolio results.

\begin{table}[!htbp]
\caption{Shared downstream LightGBM and portfolio configuration.}
\label{tab:downstream-config}
\centering
\footnotesize
\setlength{\tabcolsep}{4pt}
\renewcommand{\arraystretch}{1.10}
\begin{tabular}{>{\raggedright\arraybackslash}p{0.30\linewidth}>{\raggedright\arraybackslash}p{0.34\linewidth}>{\raggedright\arraybackslash}p{0.26\linewidth}}
\toprule
Component & Setting & Value \\
\midrule
LightGBM & Objective & Mean squared error \\
LightGBM & Learning rate; boosting cap; early stopping & $0.2$; $1000$ rounds; $50$ rounds \\
LightGBM & Column and row subsampling & $0.8879$ and $0.8789$ \\
LightGBM & $L_1$ and $L_2$ penalties & $205.6999$ and $580.9768$ \\
LightGBM & Maximum depth; leaves & $8$; $210$ \\
Portfolio & Initial account; invested fraction & $10^8$; $0.95$ \\
Portfolio & TopK-Drop rule & Hold $50$ stocks; replace up to $5$ daily \\
Portfolio & Buy, sell, and minimum-hold rules & Buy highest scores; sell lowest scores; hold threshold $1$ day \\
Portfolio & Execution and price-limit threshold & Open price; $0.095$ \\
Portfolio & Trade unit & $100$ shares \\
Portfolio & Return convention & Absolute return with a zero benchmark \\
\bottomrule
\end{tabular}
\end{table}

Predictions formed from information available through trading day $t$ are first
used for portfolio decisions at the open of trading day $t+1$.

The implementation requires Python $\geq3.12$, LightGBM $\geq4.6,<5$, and
Qlib $\geq0.9.7,<0.10$.

\subsection{Ablation-Only Configuration}

The ablations share all preceding settings unless the variant explicitly
changes a search decision. Table~\ref{tab:ablation-config} records the numerical
settings used only by the two scalarization controls. The Flow intervals were
fixed from Quality, Fertility, and AverageDistance statistics collected across
multiple Graph runs, rather than fitted to the reported ablation outcomes.

\begin{table}[!htbp]
\caption{Parameters used only by the scalarization ablations.}
\label{tab:ablation-config}
\centering
\footnotesize
\setlength{\tabcolsep}{4pt}
\renewcommand{\arraystretch}{1.12}
\begin{tabular}{>{\raggedright\arraybackslash}p{0.36\linewidth}>{\raggedright\arraybackslash}p{0.23\linewidth}>{\raggedright\arraybackslash}p{0.31\linewidth}}
\toprule
Setting & Value & Provenance \\
\midrule
Quality-scalar archive capacity & $20$ factors & Fixed ablation design \\
Flow Quality interval & $[0.44,0.71]$ & Repeated Graph-run statistics \\
Flow Fertility interval & $[0,0.004]$ & Repeated Graph-run statistics \\
Flow AverageDistance interval & $[0.60,0.86]$ & Repeated Graph-run statistics \\
Flow scalar weights & $1/3$ each & Equal-weight control \\
\bottomrule
\end{tabular}
\end{table}

\subsection{Baseline Implementations}
\label{app:baselines}
The descriptions and, where available, repository links below document the
implementation sources used for each baseline. Each baseline retains its own
factor-discovery procedure. The factors in each resulting pool are recomputed on the common data
snapshot and evaluated with the shared preprocessing, LightGBM, and backtest
protocol; reported values are not copied from the source papers or repositories.

\paragraph{Alpha158.}
We use the 158 handcrafted Qlib features as a fixed factor pool.
Implementation repository:
\url{https://github.com/microsoft/qlib}.

\paragraph{GP.}
The GP baseline adapts the genetic-programming implementation in gplearn to
the factor-mining task through a task-specific representation and search
interface. Implementation repository:
\url{https://github.com/trevorstephens/gplearn}.

\paragraph{AlphaGen.}
AlphaGen uses reinforcement learning to generate synergistic collections of
formulaic alpha factors~\citep{yu2023alphagen}. We include it as a
factor-collection baseline and evaluate its resulting pool under the common
downstream protocol. Implementation repository:
\url{https://github.com/ICT-FinD-Lab/alphagen}.

\paragraph{AlphaForge.}
AlphaForge mines formulaic alpha factors and dynamically combines them into
factor collections~\citep{shi2025alphaforge}. Its resulting pool is evaluated
with the same preprocessing, LightGBM integration, and portfolio backtest as
the other methods. Implementation repository:
\url{https://github.com/dulyhao/alphaforge}.

\paragraph{AlphaSAGE.}
AlphaSAGE performs structure-aware alpha mining with GFlowNets to encourage
diverse exploration~\citep{alphasage2026}. We use it as an additional
structure-aware search baseline and apply the common downstream evaluation
protocol to its factor pool. Implementation repository:
\url{https://github.com/BerkinChen/AlphaSAGE}.

\paragraph{AlphaAgent.}
AlphaAgent organizes factor discovery as a multi-agent workflow for hypothesis
generation, factor construction, and iterative feedback. Implementation
repository:
\url{https://github.com/RndmVariableQ/AlphaAgent/tree/legacy-main}.

\paragraph{R\&D-Agent.}
R\&D-Agent is a multi-agent framework that jointly coordinates factor research
and model development~\citep{li2025rdagentquant}. We include its factor-mining
output as a baseline and recompute downstream metrics under the common
evaluation protocol. Implementation repository:
\url{https://github.com/microsoft/rd-agent}.

\paragraph{AlphaPROBE.}
AlphaPROBE combines retrieval-guided factor proposal with graph-biased
evolution. Implementation repository:
\url{https://github.com/gta0804/AlphaPROBE}.

\paragraph{QuantaAlpha.}
QuantaAlpha performs LLM-guided evolutionary search over factor trajectories.
Implementation repository: \url{https://github.com/QuantaAlpha/QuantaAlpha}.

\paragraph{AlphaJungle and CosmosAlpha-MCTS.}
AlphaJungle~\citep{shi2026alphajungle} is an MCTS-based factor-mining method,
but we did not identify an official public implementation at the time of
submission, so we do not claim a direct reimplementation. We instead use
CosmosAlpha-MCTS as a transparent MCTS-based
proxy. It maintains multiple subtrees rather than one large tree to encourage
greater factor diversity, while sharing the factor language, evaluator, LLM
backbone, and downstream protocol with CosmosAlpha-Graph.

\subsection{CosmosAlpha-MCTS Controller}
\label{app:mcts-controller}

CosmosAlpha-MCTS uses UCT, virtual expansion, and maximum-reward backup
adapted from AlphaJungle~\citep{shi2026alphajungle}. Each iteration selects
a parent, evaluates generated children, and updates the tree and reference
archive (AlphaZoo).

\paragraph{Birth reward.}
Let $Z^-$ be the pre-admission union of current-tree and shared AlphaZoos,
deduplicated by factor ID and retaining any existing entry for $x$.
For mean daily RankIC $\mu_x$ and RankICIR $I_x$, a nonempty pool gives
\begin{align}
d_x&=1-\max_{z\in Z^-}|\rho(x,z)|,\\
p_m(x;Z^-)&=\frac{1}{|Z^-|}\sum_{z\in Z^-}\mathbb I[m_z\leq m_x],\\
r_x&=0.5p_\mu(x;Z^-)+0.25p_I(x;Z^-)+0.25p_d(x;Z^-).
\label{eq:mcts-birth-reward}
\end{align}
Reference entries retain birth metrics; $r_x$ and $d_x$ remain frozen.
For $Z^-=\emptyset$, set $d_x=1$ and
$r_x=0.5g(\mu_x)+0.25\operatorname{clip}(I_x,0,1)+0.25$, where
\begin{equation}
g(\mu)=\begin{cases}
0,&\mu\leq0,\\
[1+\exp(-80(\mu-0.025))]^{-1},&\mu>0.
\end{cases}
\end{equation}

\paragraph{UCT selection.}
Starting at the root, return a leaf or choose an unvisited child uniformly.
When all children are visited, compare UCT scores with a virtual expansion:
\begin{align}
U(s')&=B_{s'}+c\sqrt{\frac{\log N_s}{N_{s'}}},\\
V(s)&=\frac{B_s+c\sqrt{\log N_s/(1+|C(s)|)}}{1+\kappa_s}
      +\frac{0.5c(u_s/3)}{1+0.3h_s}.
\label{eq:mcts-virtual-action}
\end{align}
Here $B_s,N_s,C(s),h_s,\kappa_s$ denote maximum backed-up reward, visits,
direct children, depth, and cooling (initially zero); $c=1.6$.
The count $u_s$ records effectiveness/stability/diversity labels absent
among $C(s)$. Expand $s$ if $V(s)>\max_{s'}U(s')$; otherwise descend to the
maximizing child, resolving ties by insertion order.
Each expansion attempt independently samples a label with
$P(j\mid s)\propto\exp(-a_j)$, where
$a=(g(\mu_s),\operatorname{clip}(I_s,0,1),d_s)$.
The matched control retains label coverage but omits dimension-specific
generation instructions.

\paragraph{Backup.}
For each child, initialize $N_x=0$, $B_x=r_x$, then update every node $v$
from $x$ to the root once:
$N_v\leftarrow N_v+1$, $B_v\leftarrow\max(B_v,r_x)$.
Increment parent cooling $\kappa_s$ if $r_x<0.7r_s$, reset it if
$r_x>1.05r_s$, and otherwise retain it. Childless attempts leave visits,
backup, and cooling unchanged; their LLM calls remain charged to the Token
budget. Shared generation conditions and Token accounting are detailed in
Appendix~\ref{app:matched-budget}.

\section{Prompt Templates}
\label{app:prompt-templates}

The candidate-generation pipeline separates semantic factor design from
expression synthesis. This appendix reports the fixed instructions used to
create or repair candidate factors in the main experiments; runtime JSON
payloads are represented by placeholders and described separately below. These
are paper-facing transcriptions: escaped line breaks are rendered directly, and
implementation-only bookkeeping clauses that do not alter candidate
construction are omitted.

Table~\ref{tab:prompt-routing} records which template is used by each search
mode. Graph initialization uses the common hypothesis and factor-design
templates, whereas later Graph expansions use the parent-grounded template.
MCTS uses the common hypothesis template followed by its dedicated
factor-design template for both root generation and subsequent expansion.
Chain uses the common hypothesis and factor-design templates. All three modes
share the expression and repair stages.

\begin{table}[!htbp]
\caption{Versioned prompt routes used for candidate generation.}
\label{tab:prompt-routing}
\centering
\footnotesize
\setlength{\tabcolsep}{4pt}
\renewcommand{\arraystretch}{1.12}
\begin{tabular}{>{\raggedright\arraybackslash}p{0.25\linewidth}>{\raggedright\arraybackslash}p{0.14\linewidth}>{\raggedright\arraybackslash}p{0.25\linewidth}>{\raggedright\arraybackslash}p{0.26\linewidth}}
\toprule
Template & Version & Search stage & Structured output \\
\midrule
Hypothesis & 2.3.0 & Graph initialization and expansion; MCTS root and expansion; Chain & One testable hypothesis \\
Factor design & 2.2.0 & Graph initialization; Chain & One or two semantic designs \\
Graph factor design & 2.2.0 & Graph expansion & Parent-grounded mutations \\
MCTS factor design & 2.3.0 & MCTS root and expansion & Initial designs or parent-grounded refinements \\
Factor expression & 2.1.0 & Shared & One DSL expression per design \\
Expression repair & 2.1.0 & Shared, when validation fails & One corrected DSL expression \\
\bottomrule
\end{tabular}
\end{table}

The templates follow the version-2 prompt manifest used by the implementation.

\subsection{Runtime Substitution and Output Contracts}

The fixed messages do not enumerate a particular run's factors or numerical
search state. Instead, the renderer inserts the stage-specific JSON objects in
Table~\ref{tab:prompt-placeholders}. This separation keeps the instructions
unchanged while allowing the selected parent, measured search state, and
available factor language to vary across calls.

\begin{table}[!htbp]
\caption{Runtime placeholders appearing in the prompt templates.}
\label{tab:prompt-placeholders}
\centering
\footnotesize
\setlength{\tabcolsep}{5pt}
\renewcommand{\arraystretch}{1.12}
\begin{tabular}{>{\raggedright\arraybackslash}p{0.25\linewidth}>{\raggedright\arraybackslash}p{0.65\linewidth}}
\toprule
Placeholder & Runtime content \\
\midrule
\path{context_json} & Research goal, market, stock universe, frequency, available fields, search constraints, and bounded prior search summaries. For Graph expansion, the search constraint carries the selected parent's current guidance. \\
\path{request_json} & The current hypothesis and stage-specific request. A design request includes the universe, frequency, generation limit, and relevant parent reference. MCTS expansion additionally supplies the parent expression, scores, and target dimension. An expression request contains one semantic design and expression limits; a repair request contains the previous expression and its deterministic validation issues. \\
\path{prompt_view_json} & The runtime capability view: admissible raw fields, canonical operators and signatures, supported windows, and the bounded context permitted at that stage. \\
\bottomrule
\end{tabular}
\end{table}

Every response is parsed against a closed JSON contract: unrequested fields
are rejected. A hypothesis contains \path{statement}, \path{rationale},
\path{mechanism}, \path{testable_prediction}, \path{expected_direction},
\path{required_fields}, \path{prediction_horizon}, and \path{assumptions}. A
design response contains a \path{designs} array; each item contains exactly
\path{name}, \path{description}, \path{mechanism_mapping},
\path{signal_definition}, \path{variables}, \path{formulation},
\path{window_choices}, \path{expected_direction}, and
\path{failure_conditions}. The expression and repair stages return only
\path{proposed_expression}. The maximum array length is supplied at runtime by
\path{generation_constraints.max_designs} and is never greater than two.

\subsection{Hypothesis and Common Factor Design Templates}

The hypothesis stage asks for one falsifiable daily equity-factor hypothesis
without allowing the model to write an expression. It is used by Chain, by
both the initialization and expansion calls of Graph, and by MCTS root
generation and subsequent expansion. The paper-facing transcription below
retains the instructions that determine hypothesis content while omitting
implementation-only evidence bookkeeping.

\paragraph{Hypothesis system message.}
\begin{lstlisting}[style=prompt]
You are the CosmosAlpha Hypothesis Agent. Produce one falsifiable daily equity-factor hypothesis from the supplied research context, runtime capability catalog, and bounded structured experiment lessons. Treat prior metrics, stage outcomes, and failure codes as evidence; treat summarized lessons and next actions as prior interpretation, not new evidence. Keep the hypothesis scoped to the configured universe in the research context; do not infer or replace it with current or future constituent membership. Use only available fields. Do not write a factor expression or invent evidence. Be concise and return only the exact JSON object requested.
\end{lstlisting}

\paragraph{Hypothesis user message.}
\begin{lstlisting}[style=prompt]
Research context JSON:
${context_json}

Use context_json.universe as the research stock-pool scope. Do not replace it with another universe or use current or future constituent information.

Runtime capability and structured lesson view JSON:
${prompt_view_json}

Return exactly: statement, rationale, mechanism, testable_prediction, expected_direction, required_fields, prediction_horizon, assumptions. required_fields must contain only field names present in prompt_view_json.catalog.fields. Use prompt_view_json.lessons when present and start normally when it is empty. prediction_horizon is a string such as "1d" and assumptions is an array of strings.
\end{lstlisting}

The common factor-design stage, used by Chain and Graph initialization,
converts one hypothesis into semantic designs. It deliberately prohibits
executable syntax so that economic intent and DSL construction remain
separate.

\paragraph{Common factor-design system message.}
\begin{lstlisting}[style=prompt]
You are the CosmosAlpha Factor Design Agent. Decide what quantitative signal to construct, but do not write CosmosAlpha DSL or executable code. Every variable name and window must come from the supplied runtime capability view. Variables are raw input fields only; intermediate names such as rolling means, ratios, masks, or conditional values must not appear in variables and belong only in formulation or signal_definition. Produce one or two distinct, testable designs concisely and return only the exact JSON object requested.
\end{lstlisting}

\paragraph{Common factor-design user message.}
\begin{lstlisting}[style=prompt]
Factor design request JSON:
${request_json}

Use request_json.universe as the research stock-pool scope. Do not replace it with another universe or use current or future constituent information.

Runtime capability view JSON:
${prompt_view_json}

Return a designs array containing no more than generation_constraints.max_designs items. Each design must contain exactly: name, description, mechanism_mapping, signal_definition, variables (objects with name and role), formulation, window_choices (objects with window and rationale), expected_direction, failure_conditions. Every variables[].name must exactly match one field name in prompt_view_json.catalog.fields; use only its raw catalog name and do not list intermediate variables. variables[].role should describe the field's input role. Intermediate quantities belong only in formulation or signal_definition. Every window_choices[].window must be one of the supported_windows shown by the catalog. formulation must be mathematical or natural-language construction guidance, not DSL.
\end{lstlisting}

\subsection{Graph Parent-Grounded Mutation Template}

For a Graph expansion, the design call receives only the selected parent as its
mutation reference. Before this call, the hypothesis stage receives Graph
guidance containing the parent's expression, measured utilities, expansion
history, archive-relative gaps, and correlation-neighborhood constraints. The
resulting hypothesis and the selected parent reference then form the design
request described in Appendix~\ref{app:parent-grounded-mutation}.

\paragraph{Graph factor-design system message.}
\begin{lstlisting}[style=prompt]
You are the CosmosAlpha Graph Factor Design Agent. Produce a parent-grounded mutation of the single current parent factor described by the factor-design request JSON. Use that parent as the sole design reference. Propose one or two distinct, testable designs that make an explicit structural or signal-level change to the parent. Do not write CosmosAlpha DSL or executable code. Every variable name and window must come from the supplied runtime capability view. Variables are raw input fields only; intermediate names such as rolling means, ratios, masks, or conditional values must not appear in variables and belong only in formulation or signal_definition. Return only the exact JSON object requested.
\end{lstlisting}

\paragraph{Graph factor-design user message.}
\begin{lstlisting}[style=prompt]
Graph parent-grounded factor-design request JSON:
${request_json}

Use request_json.universe as the research stock-pool scope. Do not replace it with another universe or use current or future constituent information.

Runtime capability view JSON:
${prompt_view_json}

The request must contain only the current parent in existing_factor_refs. Treat that parent as the sole mutation reference. Return a designs array containing no more than generation_constraints.max_designs items. Each design must contain exactly: name, description, mechanism_mapping, signal_definition, variables (objects with name and role), formulation, window_choices (objects with window and rationale), expected_direction, failure_conditions. Every design must describe a mutation of the supplied parent rather than an unrelated factor. Every variables[].name must exactly match one field name in prompt_view_json.catalog.fields; use only its raw catalog name and do not list intermediate variables. variables[].role should describe the field's input role. Intermediate quantities belong only in formulation or signal_definition. Every window_choices[].window must be one of the supported_windows shown by the catalog. formulation must be mathematical or natural-language construction guidance, not DSL.
\end{lstlisting}

\subsection{MCTS Factor-Design Template}

MCTS uses the common hypothesis template followed by a dedicated factor-design
template. At root generation, its parent-specific fields are unset and the
hypothesis provides the initial design direction. During subsequent expansion,
the request supplies the selected parent expression, its available scores, and
the target search dimension; the model is instructed not to infer these
quantities from the hypothesis prose.

\paragraph{MCTS factor-design system message.}
\begin{lstlisting}[style=prompt]
You are the CosmosAlpha MCTS Factor Design Agent. Refine the supplied parent factor rather than inventing an unrelated factor. The factor-design request JSON carries the MCTS search context: parent_expression (the exact parent factor expression), parent_scores (its measured RankIC, Stability, Diversity, and Turnover when present), focus_dimension (the single search dimension this refinement must target), plus the parent metrics, target dimension, and refinement history or search guidance. Use parent_expression, focus_dimension, and parent_scores as the authoritative design basis; never reconstruct or guess them from the hypothesis text. The hypothesis mechanism is only a suggestion: if it cannot be expressed with the listed operators (for example, if it uses unsupported conditions, thresholds, or transforms), ignore that detail and refine the parent toward focus_dimension using only the listed operators. Propose one or two distinct, testable designs with an explicit relationship to the parent and a clear improvement objective for the target dimension. Do not write CosmosAlpha DSL or executable code. Every variable name and window must come from the supplied runtime capability view. Variables are raw input fields only; intermediate names such as rolling means, ratios, masks, or conditional values must not appear in variables and belong only in formulation or signal_definition. Return only the exact JSON object requested.
\end{lstlisting}

\paragraph{MCTS factor-design user message.}
\begin{lstlisting}[style=prompt]
MCTS factor-design request JSON:
${request_json}

Use request_json.universe as the research stock-pool scope. Do not replace it with another universe or use current or future constituent information.

Runtime capability view JSON:
${prompt_view_json}

Do not enumerate or exhaustively evaluate multiple alternative designs. Pick the single most promising refinement and commit to it; output it directly and keep any deliberation brief. Return a designs array containing no more than generation_constraints.max_designs items. Each design must contain exactly: name, description, mechanism_mapping, signal_definition, variables (objects with name and role), formulation, window_choices (objects with window and rationale), expected_direction, failure_conditions. Every design must refine the supplied parent factor and address the supplied target dimension; do not silently replace the parent with an unrelated idea. Every variables[].name must exactly match one field name in prompt_view_json.catalog.fields; use only its raw catalog name and do not list intermediate variables. variables[].role should describe the field's input role. Intermediate quantities belong only in formulation or signal_definition. Every window_choices[].window must be one of the supported_windows shown by the catalog. formulation must be mathematical or natural-language construction guidance, not DSL.
\end{lstlisting}

\subsection{Expression Synthesis and Repair Templates}

The expression agent translates one semantic design into the canonical DSL;
it is not allowed to redesign the signal. If deterministic validation fails,
the repair agent receives the rejected expression and concrete validation
issues and may correct only those errors. The numerical expression limits and
validity rules are specified in Appendix~\ref{app:factor-language}; the
run-specific depth cap and repair budget are reported in
Appendix~\ref{app:experiment-details}.

\paragraph{Factor-expression system message.}
\begin{lstlisting}[style=prompt]
You are the CosmosAlpha Factor Expression Agent. Translate exactly one supplied FactorDesign into one CosmosAlpha DSL expression. Use only fields, canonical operator names, signatures, and windows in the runtime capability view. Do not redesign the signal or return metadata. Be concise and return only the exact JSON object requested.
\end{lstlisting}

\paragraph{Factor-expression user message.}
\begin{lstlisting}[style=prompt]
Factor expression request JSON:
${request_json}

Runtime capability view JSON:
${prompt_view_json}

Return exactly {"proposed_expression": "..."}. The expression must be a field, number, or lowercase function call. Do not use infix arithmetic, grouping parentheses, Qlib syntax, Python, strings, attributes, keyword arguments, IDs, descriptions, or directions.
\end{lstlisting}

\paragraph{Expression-repair system message.}
\begin{lstlisting}[style=prompt]
You are the CosmosAlpha Expression Repair Agent. Correct only the deterministic expression errors supplied in the request. Preserve the FactorDesign mechanism, signal definition, direction, and formulation. Use only fields, canonical operators, signatures, and windows in the runtime capability view. Do not claim validity or change the design. Be concise and return only the exact JSON object requested.
\end{lstlisting}

\paragraph{Expression-repair user message.}
\begin{lstlisting}[style=prompt]
Expression repair request JSON:
${request_json}

Runtime capability view JSON:
${prompt_view_json}

The request contains the complete previous expression and deterministic validation issues. Repair those errors, then return exactly {"proposed_expression": "..."}. Do not return IDs, metadata, explanations, design changes, or validation status.
\end{lstlisting}

\section{Factor Language and Search Space}
\label{app:factor-language}

CosmosAlpha restricts candidate generation to a fixed domain-specific language
(DSL). Versioned field and operator catalogs define the capabilities supplied
to the LLM and accepted by the validator; a deterministic factor calculator
then executes each accepted expression. This section defines the resulting
search space: the admissible expression fields, the complete operator set, and
the syntactic and complexity constraints imposed before evaluation.

\subsection{Field Catalog}

Only the nine daily market fields in Table~\ref{tab:expression-fields} may
appear as terminals in a factor expression. Prices are post-adjusted, turnover
and returns are represented as dimensionless decimal values, and no additional
clipping or rescaling is performed by the expression engine. Missing source
observations are represented explicitly and handled according to the operator
rules below.

\begin{table}[!htbp]
\caption{Fields admissible in factor expressions. All fields are observed at
daily frequency. The domain column describes the natural data domain; entries
may also be missing.}
\label{tab:expression-fields}
\centering
\scriptsize
\setlength{\tabcolsep}{3pt}
\renewcommand{\arraystretch}{1.10}
\begin{tabular}{>{\raggedright\arraybackslash}p{0.14\linewidth}>{\centering\arraybackslash}p{0.11\linewidth}>{\raggedright\arraybackslash}p{0.43\linewidth}>{\raggedright\arraybackslash}p{0.22\linewidth}}
\toprule
DSL field & Symbol & Definition & Scale or natural domain \\
\midrule
\path{open} & $O_{i,t}$ & Opening price on trading day $t$ & Post-adjusted price; positive on eligible rows \\
\path{high} & $H_{i,t}$ & Intraday highest price & Post-adjusted price; positive on eligible rows \\
\path{low} & $L_{i,t}$ & Intraday lowest price & Post-adjusted price; positive on eligible rows \\
\path{close} & $P_{i,t}$ & Closing price & Post-adjusted price; positive on eligible rows \\
\path{volume} & $V_{i,t}$ & Daily trading volume & Source volume unit; nonnegative \\
\path{vwap} & $P^{\mathrm{VWAP}}_{i,t}$ & Volume-weighted average price & Same adjusted-price scale \\
\path{turn} & $\tau_{i,t}$ & Turnover relative to total shares outstanding & Decimal ratio; nonnegative \\
\path{free_turn} & $\tau^{\mathrm{free}}_{i,t}$ & Turnover relative to free-float shares & Decimal ratio; nonnegative \\
\path{returns} & $r_{i,t}$ & Daily return supplied by the data source & Decimal return; real-valued \\
\bottomrule
\end{tabular}
\end{table}

Table~\ref{tab:context-target-fields} separates variables needed by the
calculation and evaluation pipeline from terminals that the LLM may select.
These variables therefore cannot be inserted directly into a generated
expression. The industry code is consumed internally by
\path{indneutralize}; all five context variables are also used by the common
cross-sectional preprocessing protocol in
Table~\ref{tab:data-config}. The forward return is an evaluation target only,
as formalized in Appendix~\ref{app:single-factor-metrics}, and is never exposed
as a factor input.

\begin{table}[!htbp]
\caption{Context and target variables excluded from factor expressions. All
variables are aligned daily and constructed point-in-time.}
\label{tab:context-target-fields}
\centering
\scriptsize
\setlength{\tabcolsep}{3pt}
\renewcommand{\arraystretch}{1.10}
\begin{tabular}{>{\raggedright\arraybackslash}p{0.17\linewidth}>{\centering\arraybackslash}p{0.11\linewidth}>{\raggedright\arraybackslash}p{0.43\linewidth}>{\raggedright\arraybackslash}p{0.19\linewidth}}
\toprule
Variable & Symbol & Definition & Role and scale \\
\midrule
\path{industry_id} & $g_{i,t}$ & Point-in-time industry classification & Categorical operator and preprocessing context \\
\path{log_mkt_cap} & $m_{i,t}$ & Natural logarithm of point-in-time market capitalization & Real-valued preprocessing exposure \\
\path{ret20} & $r^{(20)}_{i,t}$ & Close-to-close return over the trailing 20 trading days through $t$ & Decimal preprocessing exposure \\
\path{turn20} & $\bar\tau^{(20)}_{i,t}$ & Mean daily turnover over the trailing 20 trading days & Nonnegative preprocessing exposure \\
\path{vol20} & $\sigma^{(20)}_{i,t}$ & Sample standard deviation of daily returns over the trailing 20 trading days & Nonnegative preprocessing exposure \\
Forward return & $y_{i,t}$ & One-trading-day forward close-to-close return & Real-valued prediction target \\
\bottomrule
\end{tabular}
\end{table}

\subsection{Operator Catalog}

The DSL contains 26 canonical operator names: 13 non-windowed and 13
time-series operators. Every operator maps one or more aligned stock--date
scalar series to another stock--date scalar series. The three supported windows
yield 52 instantiated primitives---13 non-windowed and $13\times3$ windowed
variants---before recursive composition. In
Table~\ref{tab:operator-catalog}, ``Comm.'' marks
operators whose inputs are reordered only when CosmosAlpha constructs Graph
identities and structural representations; ordinary expression serialization
preserves the written argument order.

\begingroup
\scriptsize
\setlength{\tabcolsep}{3pt}
\renewcommand{\arraystretch}{1.08}
\setlength{\LTleft}{0pt}
\setlength{\LTright}{0pt}
\setlength{\LTcapwidth}{\linewidth}
\begin{longtable}{@{}>{\raggedright\arraybackslash}p{0.21\linewidth}>{\raggedright\arraybackslash}p{0.14\linewidth}>{\raggedright\arraybackslash}p{0.47\linewidth}>{\centering\arraybackslash}p{0.08\linewidth}@{}}
\caption{Complete operator catalog. Here $\operatorname{rank}_t$ is the
same-day cross-sectional percentile rank over eligible finite stocks, with
average ranks for ties; $\varepsilon=10^{-12}$; and every window argument is
an integer $w\in\{5,10,20\}$ measured in trading days.}
\label{tab:operator-catalog}\\
\toprule
Signature & Scope & Definition and numerical behavior & Comm. \\
\midrule
\endfirsthead
\multicolumn{4}{@{}l}{\textit{Table~\thetable\ continued.}}\\
\toprule
Signature & Scope & Definition and numerical behavior & Comm. \\
\midrule
\endhead
\midrule
\multicolumn{4}{r@{}}{\textit{Continued on next page}}\\
\endfoot
\bottomrule
\endlastfoot
\multicolumn{4}{@{}l}{\textit{Non-windowed operators}}\\*
\midrule
\path{add(x,y)} & Elementwise & $x+y$ for finite inputs & Yes \\
\path{sub(x,y)} & Elementwise & $x-y$ for finite inputs & No \\
\path{mul(x,y)} & Elementwise & $xy$ for finite inputs & Yes \\
\path{div(x,y)} & Elementwise & $x/y$; missing when $|y|\leq\varepsilon$ & No \\
\path{abs(x)} & Elementwise & $|x|$ & No \\
\path{sqrt(x)} & Elementwise & $\sqrt{x}$ for $x\geq0$; otherwise missing & No \\
\path{log(x)} & Elementwise & $\log x$ for $x>0$; otherwise missing & No \\
\path{inv(x)} & Elementwise & $1/x$; missing when $|x|\leq\varepsilon$ & No \\
\path{rank(x)} & Cross-sectional & $\operatorname{rank}_t(x)\in(0,1]$ & No \\
\path{rank_sub(x,y)} & Cross-sectional & $\operatorname{rank}_t(x)-\operatorname{rank}_t(y)$ & No \\
\path{rank_div(x,y)} & Cross-sectional & $\operatorname{rank}_t(x)/\operatorname{rank}_t(y)$ with protected division & No \\
\path{sigmoid(x)} & Elementwise & $(1+\exp(-x))^{-1}$ & No \\
\path{indneutralize(x)} & Industry group & $x$ minus its same-day mean over eligible finite stocks in industry $g_{i,t}$ & No \\
\addlinespace
\multicolumn{4}{@{}l}{\textit{Time-series operators}}\\*
\midrule
\path{delay(x,w)} & Time series & Value $x_{i,t-w}$ from exactly $w$ trading rows earlier & No \\
\path{delta(x,w)} & Time series & $x_{i,t}-x_{i,t-w}$ & No \\
\path{corr(x,y,w)} & Time series & Rolling Pearson correlation on pairwise finite observations & Yes \\
\path{cov(x,y,w)} & Time series & Rolling sample covariance on pairwise finite observations & Yes \\
\path{decay_linear(x,w)} & Time series & Linearly weighted mean with weights $1,\ldots,w$, largest for the most recent observation & No \\
\path{ts_min(x,w)} & Time series & Rolling minimum over finite observations & No \\
\path{ts_max(x,w)} & Time series & Rolling maximum over finite observations & No \\
\path{ts_argmin(x,w)} & Time series & One-based position of the first rolling minimum within the window & No \\
\path{ts_argmax(x,w)} & Time series & One-based position of the first rolling maximum within the window & No \\
\path{ts_rank(x,w)} & Time series & Percentile rank of $x_{i,t}$ among finite window observations, with averaged ties & No \\
\path{ts_sum(x,w)} & Time series & Rolling sum over finite observations & No \\
\path{ts_stddev(x,w)} & Time series & Rolling sample standard deviation ($\mathrm{ddof}=1$) & No \\
\path{ts_zscore(x,w)} & Time series & $(x_{i,t}-\bar x_{i,t,w})/s_{i,t,w}$ using the rolling sample deviation & No \\
\end{longtable}
\endgroup

Elementwise operators propagate missing inputs. Window reducers require at
least $\lceil2w/3\rceil$ finite observations, while binary rolling statistics
apply the same rule to pairwise finite observations; \path{delay} and
\path{delta} instead require the relevant current and lagged values. The
linear-decay weights are renormalized over the available observations. Rolling
standard deviations and $z$-scores are missing when their denominator is below
$\varepsilon$, as are correlations with a degenerate input. Any remaining
non-finite result is stored as missing before evaluation.

\subsection{Grammar, Validation, and Complexity Bounds}

An expression is a field, a numeric constant, or a prefix function call whose
arguments are themselves expressions:
\begin{equation}
e ::= f \mid c \mid o(e_1,\ldots,e_k)
       \mid o_w(e_1,\ldots,e_k,w),
\qquad
f\in\mathcal F,\quad w\in\{5,10,20\},
\label{eq:factor-dsl-grammar}
\end{equation}
where $\mathcal F$ is the expression-field set in
Table~\ref{tab:expression-fields}. Function and field identifiers are
lowercase, and numeric literals use ordinary integer or decimal notation.
Arithmetic must use the registered prefix functions; infix operators,
standalone grouping parentheses, scientific notation, free-form Python, Qlib
syntax, strings, attributes, and keyword arguments are outside the language.
Constants may occur within an expression, but every complete factor must
reference at least one admissible field.

Table~\ref{tab:expression-constraints} summarizes the hard checks applied
before factor calculation. Operators may be nested without a separate
operator-pair restriction, subject to the AST-depth and node-count caps. The
validator also checks each function's exact arity, its final window literal,
the availability of any required context, the daily frequency, and backend
support. Consequently, a syntactically valid expression can still be rejected
when it lies outside the registered search space.

\begin{table}[!htbp]
\caption{Expression-space and validation constraints. Method-specific search
budgets and repair limits are reported in
Appendix~\ref{app:experiment-details}.}
\label{tab:expression-constraints}
\centering
\scriptsize
\setlength{\tabcolsep}{4pt}
\renewcommand{\arraystretch}{1.10}
\begin{tabular}{>{\raggedright\arraybackslash}p{0.28\linewidth}>{\raggedright\arraybackslash}p{0.62\linewidth}}
\toprule
Constraint & Enforced rule \\
\midrule
Data frequency & Daily (\path{1d}) for every field and operator \\
Allowed windows & Final integer literal in $\{5,10,20\}$ trading days \\
Expression length & At most 512 characters at the parser boundary \\
AST depth & Root counted at depth 1; the run-specific cap appears in Table~\ref{tab:search-config} \\
Operator count & At most 32 call nodes \\
Field count & At most 6 distinct expression fields and at least 1 field \\
Temporal access & Only contemporaneous or backward-looking fields and windows; target and future-return fields are unavailable \\
Catalog compliance & Registered lowercase fields and operators, exact arity, available context, and supported backend \\
Missing values & Undefined domains, insufficient rolling observations, and non-finite outputs remain missing \\
\bottomrule
\end{tabular}
\end{table}

The syntax normalizer produces deterministic lowercase prefix text but does
not perform algebraic simplification or reorder operands. Graph identity and
structural extraction then apply the narrower commutative rule indicated in
Table~\ref{tab:operator-catalog}: the two inputs of \path{add}, \path{mul},
\path{corr}, and \path{cov} are sorted, while
the window of a rolling operator remains fixed. Duplicate handling after this
normalization is described in Appendix~\ref{app:factor-identity}.

\section{Qualitative Factor Examples}
\label{app:factor-examples}

This section makes the output of CosmosAlpha concrete through four factors
from a CSI~300 Graph diagnostic run with 185 nodes. The examples are descriptive
rather than an additional performance comparison: each case reports its exact
DSL expression, plain-language operation, cautious economic interpretation, and
training-period single-factor statistics. E1--E4 are case labels; their
six-character factor IDs match the corresponding nodes in the following lineage
figure.

\subsection{Case Selection and Reporting Protocol}

We select cases only from technically valid, unique factors in the final
Quality Pareto Archive and do not inspect validation- or test-period outcomes.
E1 is the factor with the largest absolute mean RankIC; E2 is the remaining
factor with the smallest RankIC standard deviation; and E3 is the remaining
factor with the largest direction-aligned Top-20\% return. E4 is the remaining
archive factor with the largest final-round AverageDistance. If one factor
occupies more than one quality extreme, the next distinct factor under the
corresponding ordering is used. This rule represents the three intrinsic-quality
objectives and a comparatively non-redundant point, rather than selecting four
cases by one scalar score or by held-out performance.

Table~\ref{tab:qualitative-factor-metrics} records each case's graph ID and
generation round and depth. The first three statistics follow
Appendix~\ref{app:single-factor-metrics}; the Top-20 return is direction-aligned
and shown in basis points per trading day. AverageDistance is the final-round
value defined in Eq.~\ref{eq:average-distance}. These are search-time diagnostics
on the training split, not the integrated LightGBM test metrics in
Tables~\ref{tab:main-csi300} and~\ref{tab:main-csi500}.

\begin{table}[!htbp]
\caption{Training-period diagnostics for the representative CSI~300 Graph
factors. Direction-aligned Top-20 return is reported in basis points (bp) per trading day.}
\label{tab:qualitative-factor-metrics}
\centering
\scriptsize
\setlength{\tabcolsep}{3.5pt}
\renewcommand{\arraystretch}{1.10}
\begin{tabular}{>{\centering\arraybackslash}p{0.07\linewidth}>{\centering\arraybackslash}p{0.11\linewidth}>{\centering\arraybackslash}p{0.10\linewidth}>{\centering\arraybackslash}p{0.14\linewidth}>{\centering\arraybackslash}p{0.14\linewidth}>{\centering\arraybackslash}p{0.14\linewidth}>{\centering\arraybackslash}p{0.14\linewidth}}
\toprule
Case & Graph ID & Round/$d$ &
\makecell{$|\overline{\mathrm{RankIC}}|$\\$\uparrow$} &
\makecell{$\sigma(\mathrm{RankIC})$\\$\downarrow$} &
\makecell{Aligned Top-20\\(bp/day) $\uparrow$} &
\makecell{AverageDistance\\$\uparrow$} \\
\midrule
E1 & \texttt{63d83d} & 18/5 & 0.058165 & 0.073829 & 13.10 & 0.5092 \\
E2 & \texttt{835f66} & 20/7 & 0.005304 & 0.057119 & 4.93 & 0.8182 \\
E3 & \texttt{1b4d31} & 21/4 & 0.054396 & 0.072582 & 12.79 & 0.5040 \\
E4 & \texttt{5a02af} & 19/7 & 0.017627 & 0.063906 & 3.85 & 0.8535 \\
\bottomrule
\end{tabular}
\end{table}

\newpage
\subsection{Representative Factors}

\begin{samepage}
\paragraph{E1: RankIC-oriented case.}
\emph{Expression:} \path{decay_linear(div(sub(close, vwap), sub(high, low)), 5)}.
\emph{Reading:} For each stock, the expression scales the close--VWAP gap by
the day's high--low range, then applies a five-day linearly weighted mean that
gives more weight to recent observations.
\emph{Interpretation:} Its signed training mean RankIC is $-0.0582$; the
direction-aligned Top-20 statistic therefore reverses the raw factor orientation
as specified in Appendix~\ref{app:single-factor-metrics}. The expression can be
read as a smoothed, range-normalized displacement from VWAP, while the observed
association alone does not establish an economic mechanism.
\end{samepage}

\begin{samepage}
\paragraph{E2: stability-oriented case.}
\emph{Expression:} \path{rank_sub(ts_zscore(free_turn, 5), ts_zscore(div(sub(high, low), open), 5))}.
\emph{Reading:} It computes five-day time-series $z$-scores for free-float
turnover and the high--low range relative to the opening price, then subtracts
their same-day cross-sectional ranks.
\emph{Interpretation:} E2 has the smallest RankIC standard deviation among the
remaining archive cases ($0.0571$), although its absolute mean RankIC is only
$0.0053$. It illustrates the stability criterion; the low variation does not
by itself imply a strong mean cross-sectional association.
\end{samepage}

\begin{samepage}
\paragraph{E3: Top-20-return-oriented case.}
\emph{Expression:} \path{mul(decay_linear(div(sub(close, vwap), sub(high, low)), 5), rank(free_turn))}.
\emph{Reading:} This multiplies the five-day smoothed, range-normalized
close--VWAP gap by the same-day cross-sectional rank of free-float turnover.
\emph{Interpretation:} Among the remaining archive cases, it has the largest
direction-aligned Top-20 training return (12.79 bp per trading day). This is a
single-factor search diagnostic and does not establish downstream portfolio
performance or explain the source of the association.
\end{samepage}

\begin{samepage}
\paragraph{E4: diversity-oriented case.}
\emph{Expression:} \path{mul(div(close, open), sub(ts_rank(free_turn, 10), delay(ts_rank(free_turn, 10), 5)))}.
\emph{Reading:} It multiplies the close-to-open price ratio by the change in
the ten-day time-series rank of free-float turnover relative to five trading
days earlier.
\emph{Interpretation:} E4 has the largest final-round AverageDistance among
the remaining archive cases ($0.8535$), indicating separation under the
paper's structure and cross-sectional exposure-correlation distances. This is a diversity
diagnostic, not evidence of predictive strength or an economic mechanism.
\end{samepage}

Taken together, the four cases illustrate distinct selection criteria in the
intrinsic Quality Pareto Archive. The next subsection lists all factors in the
terminal search Pareto front; the subsequent lineage figure shows their
generation paths and front assignments.

\subsection{Members of the First Search Pareto Front}
The catalog in Table~\ref{tab:search-front1-factors} lists all 44 nodes in the
terminal first search Pareto front, ordered by generation round and graph ID.
This front is formed by non-dominated sorting of propagated Quality, Fertility,
and Diversity utilities. It differs from the intrinsic Quality Pareto Archive
used to select E1--E4, although seven nodes belong to both sets. The six-character
IDs match those in the lineage figure; $d$ denotes generation depth. The two
initial factors are marked \emph{Init.} and have generation depth $d=0$.

\begingroup
\scriptsize
\setlength{\tabcolsep}{3pt}
\renewcommand{\arraystretch}{1.08}
\setlength{\LTleft}{0pt}
\setlength{\LTright}{0pt}
\setlength{\LTcapwidth}{\linewidth}
\begin{longtable}{@{}>{\centering\arraybackslash\ttfamily}p{0.11\linewidth}>{\centering\arraybackslash}p{0.15\linewidth}>{\raggedright\arraybackslash}p{0.68\linewidth}@{}}
\caption{Canonical expressions for all 44 factors in terminal search Pareto front 1. The ID is the six-character prefix used in the lineage figure; Round is the generation round and $d$ is generation depth.}\\
\label{tab:search-front1-factors}\\
\toprule
ID & Round/$d$ & Canonical DSL expression \\
\midrule
\endfirsthead
\caption[]{Canonical expressions for terminal search Pareto front 1 (continued).}\\
\toprule
ID & Round/$d$ & Canonical DSL expression \\
\midrule
\endhead
\midrule
\multicolumn{3}{r}{Continued on the next page}\\
\endfoot
\bottomrule
\endlastfoot
a6f90c & Init./0 & \path{mul(mul(sub(close, low), inv(sub(high, low))), ts_rank(volume, 20))} \\
ed54a5 & Init./0 & \path{rank(mul(sub(div(close, vwap), 1), div(turn, div(ts_sum(turn, 20), 20))))} \\
65260c & 0/1 & \path{mul(div(turn, div(ts_sum(turn, 20), 20)), div(sub(vwap, close), sub(high, low)))} \\
4060a1 & 1/1 & \path{rank(mul(div(sub(close, vwap), sub(high, low)), ts_rank(volume, 20)))} \\
19f0cc & 2/3 & \path{mul(div(sub(close, open), open), ts_zscore(free_turn, 20))} \\
f71706 & 2/2 & \path{mul(ts_zscore(div(sub(close, vwap), sub(high, low)), 10), div(volume, div(ts_sum(volume, 20), 20)))} \\
4b967c & 3/1 & \path{mul(div(sub(close, low), sub(high, low)), div(sub(turn, div(ts_sum(turn, 20), 20)), ts_stddev(turn, 20)))} \\
c192ad & 3/1 & \path{mul(div(sub(close, vwap), vwap), ts_zscore(free_turn, 20))} \\
cc0000 & 3/4 & \path{mul(sub(div(close, vwap), 1), ts_rank(free_turn, 20))} \\
fa679f & 3/1 & \path{rank(mul(rank(div(sub(vwap, low), sub(high, low))), rank(div(free_turn, div(ts_sum(free_turn, 20), 20)))))} \\
dcbd43 & 4/2 & \path{mul(delta(div(sub(close, vwap), sub(high, low)), 5), ts_rank(volume, 20))} \\
0cdf59 & 5/1 & \path{mul(rank(div(sub(close, vwap), sub(high, low))), rank(sigmoid(ts_zscore(turn, 20))))} \\
a2fda7 & 5/2 & \path{mul(sub(0, rank(sub(div(close, vwap), 1))), rank(ts_zscore(free_turn, 5)))} \\
c9d72b & 5/2 & \path{mul(sub(0, rank(ts_zscore(sub(div(close, vwap), 1), 10))), rank(free_turn))} \\
fc4167 & 5/1 & \path{mul(rank(sub(div(close, vwap), 1)), rank(sigmoid(ts_zscore(free_turn, 20))))} \\
09a17d & 6/4 & \path{rank(mul(div(sub(vwap, low), sub(high, low)), corr(div(sub(vwap, low), sub(high, low)), ts_zscore(turn, 20), 5)))} \\
8419c4 & 6/4 & \path{mul(div(sub(close, vwap), ts_stddev(close, 20)), log(add(1, div(volume, decay_linear(volume, 10)))))} \\
c4ae40 & 6/2 & \path{mul(div(sub(close, vwap), sub(ts_max(high, 10), ts_min(low, 10))), ts_rank(volume, 10))} \\
f369c2 & 6/2 & \path{mul(ts_sum(div(sub(close, vwap), sub(high, low)), 5), ts_rank(volume, 5))} \\
35f75c & 7/1 & \path{mul(sub(turn, delay(turn, 5)), div(sub(close, vwap), vwap))} \\
5ecd5e & 8/1 & \path{rank(mul(sub(div(close, vwap), 1), div(mul(turn, 20), ts_sum(turn, 20))))} \\
6cb762 & 8/1 & \path{rank(mul(mul(sub(div(close, vwap), 1), div(turn, div(ts_sum(turn, 20), 20))), add(1, div(sub(high, low), close))))} \\
1340e2 & 10/3 & \path{decay_linear(mul(free_turn, div(sub(close, vwap), sub(high, low))), 5)} \\
3895f5 & 13/4 & \path{mul(div(sub(sub(mul(2, close), high), low), sub(high, low)), corr(div(sub(close, low), sub(high, low)), free_turn, 10))} \\
1185be & 14/4 & \path{decay_linear(mul(rank(free_turn), ts_zscore(div(sub(close, vwap), sub(high, low)), 20)), 10)} \\
85f0b6 & 14/4 & \path{decay_linear(mul(rank(free_turn), rank(div(sub(close, vwap), sub(high, low)))), 5)} \\
3ae17f & 17/5 & \path{mul(div(div(sub(close, vwap), vwap), ts_stddev(div(sub(close, vwap), vwap), 10)), turn)} \\
427d2c & 17/5 & \path{rank(mul(ts_rank(free_turn, 10), rank(sub(close, open))))} \\
fa88b4 & 17/5 & \path{mul(div(sub(close, vwap), sub(high, low)), turn)} \\
63d83d & 18/5 & \path{decay_linear(div(sub(close, vwap), sub(high, low)), 5)} \\
c81478 & 18/6 & \path{mul(rank(div(close, open)), delta(ts_rank(free_turn, 10), 5))} \\
12e0bc & 19/3 & \path{mul(ts_zscore(sub(close, vwap), 20), ts_rank(volume, 10))} \\
5a02af & 19/7 & \path{mul(div(close, open), sub(ts_rank(free_turn, 10), delay(ts_rank(free_turn, 10), 5)))} \\
751385 & 19/5 & \path{add(rank(div(sub(vwap, low), sub(high, low))), rank(cov(div(sub(vwap, low), sub(high, low)), ts_zscore(turn, 20), 10)))} \\
7aabfd & 19/5 & \path{add(rank(div(sub(vwap, low), sub(high, low))), rank(corr(div(sub(vwap, low), sub(high, low)), delta(ts_zscore(free_turn, 20), 5), 5)))} \\
bb8068 & 19/3 & \path{mul(div(sub(close, vwap), sub(ts_max(high, 10), ts_min(low, 10))), rank(volume))} \\
5b9526 & 20/5 & \path{mul(mul(div(div(sub(vwap, close), sub(high, low)), abs(div(sub(vwap, close), sub(high, low)))), rank(abs(div(sub(vwap, close), sub(high, low))))), rank(turn))} \\
902232 & 20/8 & \path{mul(log(div(close, open)), sub(ts_rank(free_turn, 10), delay(ts_rank(free_turn, 10), 5)))} \\
d819dd & 20/8 & \path{mul(sub(div(close, open), 1), delta(ts_zscore(free_turn, 10), 5))} \\
8eb196 & 22/9 & \path{mul(rank(sub(div(close, open), 1)), rank(delta(free_turn, 5)))} \\
5cc5bf & 23/3 & \path{rank(mul(ts_zscore(div(sub(vwap, close), sub(high, low)), 10), div(volume, delay(volume, 5))))} \\
88860d & 23/3 & \path{rank(mul(div(sub(vwap, close), sub(high, low)), div(free_turn, delay(free_turn, 5))))} \\
88e162 & 24/2 & \path{mul(sub(div(close, vwap), 1), ts_rank(turn, 20))} \\
a5dfab & 24/9 & \path{mul(div(sub(close, vwap), vwap), ts_rank(free_turn, 10))} \\
\end{longtable}
\endgroup

The catalog provides the expressions for the full first front, while the four
cases above give more detailed readings of selected archive members. The
following lineage figure places both sets within the same generation graph.

\section{Factor Lineage and Search Pareto Fronts}
\label{app:factor-lineage}

Figure~\ref{fig:factor-lineage} shows the generation lineage for the CSI~300
Graph diagnostic run used in Appendix~\ref{app:factor-examples}. This run is
separate from Table~\ref{tab:main-csi300}; its 185 nodes comprise two initial
factors and 183 subsequently generated unique valid factors. The approximately
150-factor target was not imposed as a hard cap. Each node's fill color
identifies its terminal search Pareto front, computed from the propagated Quality,
Fertility, and Diversity utilities. Front~1 contains the non-dominated nodes;
subsequent fronts are obtained by repeatedly removing the preceding front before
sorting the remaining nodes. Thus, the fill colors encode a node's rank in the
three-objective search ordering, not its Quality alone. Counts beside F1--F7 in
the legend give the number of nodes in each front.

Edges connect generated factors to their parents. Node labels show a shortened
factor ID, generation depth ($d$), and front index.

\begin{figure}[!htbp]
\centering
\begin{overpic}[height=0.88\textheight,keepaspectratio]{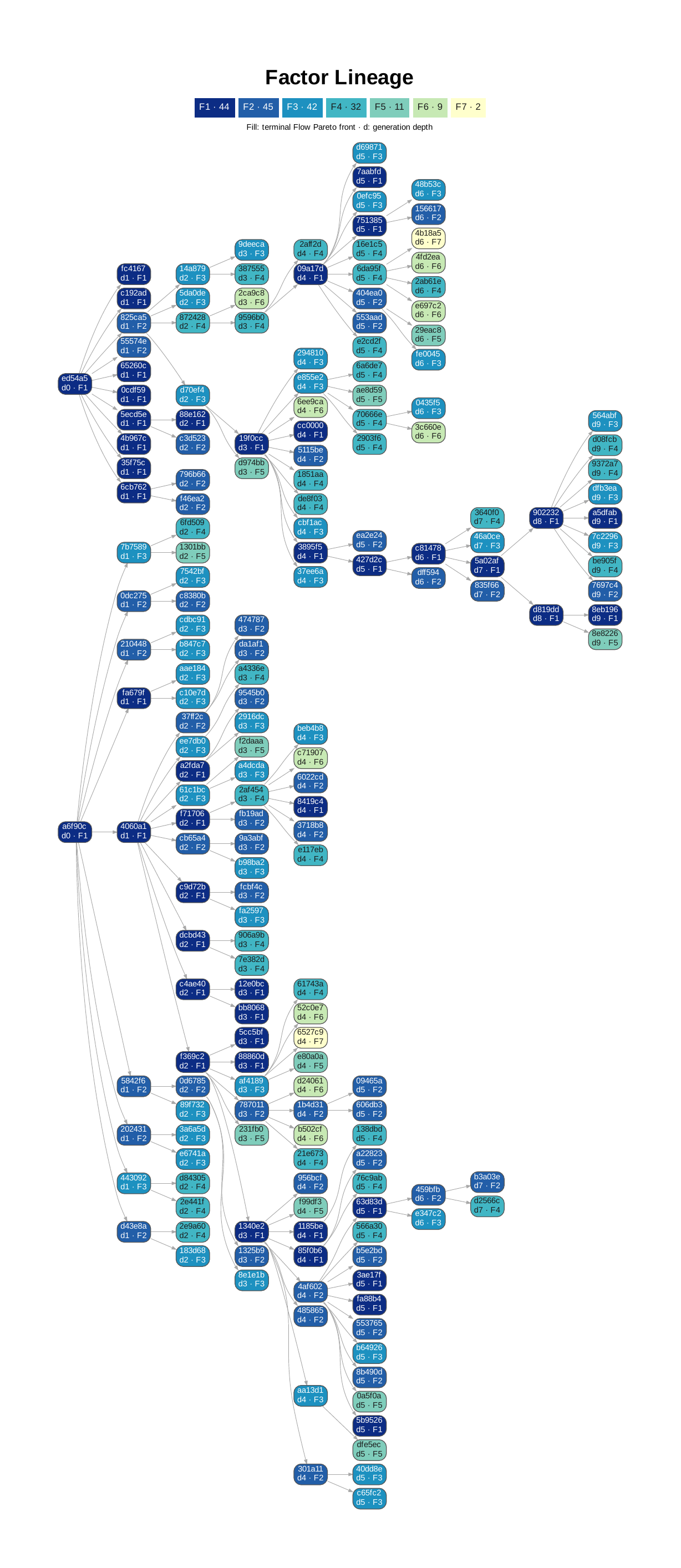}
  \put(13.0,91.25){\colorbox{white}{\rule{0pt}{6pt}\makebox[98pt]{}}}
  \put(13.4,91.70){\makebox[94pt][c]{\sffamily\fontsize{3.2}{3.6}\selectfont
  Fill: terminal search Pareto front \textperiodcentered{} \textit{d}: generation depth}}
\end{overpic}
\caption{Factor lineage and terminal search Pareto-front assignments for the
CSI~300 Graph diagnostic run with 185 nodes. Node fill color marks the front;
the legend reports the number of nodes in each front. Edges show direct
generation links, and each node lists its shortened ID, generation depth $d$,
and front index.}
\label{fig:factor-lineage}
\end{figure}

\section{Ablation Details}
\label{app:ablation-semantics}
The ablation variants follow the common CSI 300 data split, LLM configuration,
prompt protocol, search budget, and downstream LightGBM evaluation. The two
scalarization variants modify separate decisions in the search process.

\paragraph{Intrinsic-quality scalarization.}
The control uses the same raw intrinsic-quality coordinates
\[
q_i=\left(|\overline{\mathrm{RankIC}}_i|,
-\sigma(\mathrm{RankIC}_i),R_i^{\mathrm{Top20}}\right).
\]
Quality scalarization retains the fixed normalization in
Appendix~\ref{app:quality-details} and averages its three coordinates:
\begin{equation}
S_i=\frac{\widetilde q_i^{\mathrm{IC}}+
\widetilde q_i^{\mathrm{std}}+\widetilde q_i^{\mathrm{ret}}}{3}.
\label{eq:ablation-self-score}
\end{equation}
This score is also used as the node's SelfQuality in the full model. The
ablation additionally retains the highest-scoring factors up to the fixed
archive capacity in Table~\ref{tab:ablation-config}, instead of retaining the
Pareto first front. Search Pareto selection remains in place. This archive
change affects two subsequent quantities.

\emph{Quality reference set.} AverageDistance measures structure and
correlation distance to a high-quality factor set from the current run. In the
full model, this set is the intrinsic-quality Pareto archive. Under Quality
scalarization, the same role is filled by the highest-scoring factors under
$S_i$. Thus, the ablation changes which factors provide the structure and
correlation references, while the pairwise-correlation formula and graph-edge
construction remain unchanged. The reference set is fixed for each round and
shared by all nodes in that round, and is built only from the current run's
factor archive.

\emph{Fertility.} In the full model, a child's contribution is its increment
to the intrinsic-quality archive's hypervolume. Quality scalarization instead
measures its gain over the best scalar score in the archive frozen before the
round:
\begin{equation}
\Delta S(c)=\max\left(0,S(c)-\max_{a\in\mathcal A_r^-}S(a)\right),
\label{eq:ablation-delta-s}
\end{equation}
where $\mathcal A_r^-$ is the pre-round scalar archive. Each child is evaluated
against that same archive, so siblings do not affect one another's gain. An
expansion attempt takes the largest gain among its valid children, and parent
Fertility remains the Top-3 mean of historical attempt outcomes. The scalar
archive also supplies the factor-quality gaps passed to the generation agent.

\paragraph{Flow scalarization.}
Flow scalarization retains the intrinsic-quality Pareto archive, hypervolume
gain, and Graph Flow propagation. It replaces Pareto front and crowding-based
parent ranking with an equal-weight score over the propagated utilities
$(Q,F,D)$. For utility $m\in\{Q,F,D\}$, let $(\ell_m,u_m)$ be its fixed
normalization interval from Table~\ref{tab:ablation-config}. The ablation uses
\begin{equation}
z_{m,i}=\operatorname{clip}\!\left(
\frac{F_{m,i}-\ell_m}{u_m-\ell_m},0,1\right),
\label{eq:ablation-flow-normalization}
\end{equation}
where $F_{m,i}$ is node $i$'s propagated value for utility $m$. The intervals
were estimated from multiple Graph runs and fixed before the ablation.
Eligible parents are then ranked by
\begin{equation}
s_i^{\mathrm{flow}}=\frac{z_{Q,i}+z_{F,i}+z_{D,i}}{3}.
\label{eq:ablation-flow-score}
\end{equation}
The fixed intervals prevent the different numeric scales of the three
utilities from determining the equal-weight sum. Eligible parents are ranked
by descending $s_i^{\mathrm{flow}}$; eligibility, search budget, and the number
of selected parents follow the full model.

\paragraph{Flow, utility, and relation controls.}
Setting $\gamma=0$ removes neighbor propagation while retaining node
utilities. Utility ablations restrict search Pareto selection to $(Q,D)$,
$(Q,F)$, or $(Q)$ for the Fertility, Diversity, and Quality-only rows,
respectively. Removing Structure or Correlation excludes that relation from
both Graph Flow and AverageDistance; removing Generation excludes it from
Graph Flow. The remaining Flow relation weights are renormalized.
Appendix~\ref{app:relation-role-controls} adds single-role controls that remove
Structure or Correlation from only propagation or only AverageDistance.

\section{LLM Sensitivity Study}
\label{app:llm-sensitivity}

This appendix compares DeepSeek-V4-Flash, GPT-5.6-Sol, and Kimi K3 under the
Graph, MCTS, and Chain modes of CosmosAlpha on CSI 300. All entries use the
same factor-evaluation and downstream backtest conventions as the main tables.

\begin{table}[!htbp]
\caption{LLM and search-mode sensitivity on CSI 300 under the main factor-pool integration protocol. ARR and signed MDD are percentages; higher values are better for every displayed metric because a signed MDD closer to zero denotes a smaller drawdown.}
\label{tab:appendix-llm}
\centering
\footnotesize
\setlength{\tabcolsep}{4pt}
\renewcommand{\arraystretch}{1.15}
\begin{tabular*}{\linewidth}{@{\extracolsep{\fill}}llrrrrrrr@{}}
\toprule
\multirow{2}{*}{Mode} & \multirow{2}{*}{LLM} & \multicolumn{4}{c}{Predictive} & \multicolumn{3}{c}{Portfolio} \\
\cmidrule(lr){3-6} \cmidrule(lr){7-9}
& & IC & ICIR & RankIC & RankICIR & ARR & IR & MDD \\
\midrule
\multirow{3}{*}{Chain} & DeepSeek-V4-Flash & 0.0241 & 0.2731 & 0.0373 & 0.4260 & 3.94 & 0.2223 & -35.33 \\
& GPT-5.6-Sol & 0.0280 & 0.2892 & 0.0411 & 0.4519 & 4.16 & 0.2599 & -29.11 \\
& Kimi K3 & 0.0299 & 0.3156 & 0.0401 & 0.3932 & 4.28 & 0.2810 & -33.60 \\
\cmidrule(lr){2-9}
\multirow{3}{*}{MCTS} & DeepSeek-V4-Flash & 0.0281 & 0.3238 & 0.0409 & 0.5145 & 5.68 & 0.3296 & -31.56 \\
& GPT-5.6-Sol & 0.0259 & 0.2868 & 0.0419 & 0.4590 & 5.95 & 0.3361 & -33.81 \\
& Kimi K3 & 0.0289 & 0.3025 & 0.0412 & 0.5143 & 4.23 & 0.2502 & -34.99 \\
\cmidrule(lr){2-9}
\multirow{3}{*}{Graph} & DeepSeek-V4-Flash & 0.0324 & 0.3441 & 0.0494 & 0.5670 & 8.44 & 0.5332 & -25.07 \\
& GPT-5.6-Sol & 0.0294 & 0.3379 & 0.0431 & 0.5119 & 7.51 & 0.4358 & -26.59 \\
& Kimi K3 & 0.0324 & 0.3818 & 0.0463 & 0.5909 & 6.11 & 0.3519 & -28.91 \\
\bottomrule
\end{tabular*}
\end{table}

Across all three backbone models, Graph delivers the strongest overall
performance and leads all seven metrics for each backbone. The portfolio results particularly
favor DeepSeek-V4-Flash in the Graph setting: it leads ARR, IR, and signed MDD
for Graph. For MCTS, GPT-5.6-Sol has slightly higher ARR and IR than
DeepSeek-V4-Flash (5.95 versus 5.68 and 0.3361 versus 0.3296), whereas
DeepSeek-V4-Flash has the better signed MDD (-31.56\% versus -33.81\%). Across
each search mode, the three backbone models remain relatively
close on Predictive Performance, whereas the differences between Graph, MCTS,
and Chain are more pronounced, especially for Portfolio Performance.

\clearpage
\section{Independent-Run Stability Study}
\label{app:run-stability}

This appendix evaluates the stability of CosmosAlpha-Graph,
CosmosAlpha-MCTS, and CosmosAlpha-Chain with DeepSeek-V4-Flash on CSI 300
across three independent LLM runs. All three methods are reported in one table
under the same factor-pool integration protocol as the main results. The Mean
and Std. rows are computed across the three runs for each method according to
Appendix~\ref{app:repeated-run-metrics}.
The data split and experimental configuration remain fixed across repetitions,
so the variation reflects stochasticity in the factors generated by the LLM.

\begin{table}[!htbp]
\caption{Three-run stability of CosmosAlpha-Graph, CosmosAlpha-MCTS, and CosmosAlpha-Chain with DeepSeek-V4-Flash on CSI 300 under the main integration protocol. The table uses the same predictive and portfolio metric groups and metric directions as Tables~\ref{tab:main-csi300} and~\ref{tab:main-csi500}. ARR and signed MDD are percentages. The Mean and Std. rows are computed across the three independent runs for each method; Std. uses the population standard deviation.}
\label{tab:appendix-run}
\centering
\footnotesize
\setlength{\tabcolsep}{4pt}
\renewcommand{\arraystretch}{1.15}
\begin{tabular*}{\linewidth}{@{\extracolsep{\fill}}llrrrrrrr@{}}
\toprule
\multirow{2}{*}{Mode} & \multirow{2}{*}{Run} & \multicolumn{4}{c}{Predictive} & \multicolumn{3}{c}{Portfolio} \\
\cmidrule(lr){3-6} \cmidrule(lr){7-9}
& & IC & ICIR & RankIC & RankICIR & ARR & IR & MDD \\
\midrule
\multirow{5}{*}{Chain} & Run 1 & 0.0230 & 0.2606 & 0.0369 & 0.4087 & 3.49 & 0.1839 & -39.27 \\
& Run 2 & 0.0234 & 0.2732 & 0.0371 & 0.4261 & 4.06 & 0.2399 & -33.13 \\
& Run 3 & 0.0242 & 0.2912 & 0.0352 & 0.4332 & 4.48 & 0.2631 & -36.60 \\
\cmidrule(lr){2-9}
& Mean & 0.0235 & 0.2750 & 0.0364 & 0.4227 & 4.01 & 0.2290 & -36.33 \\
& Std. & 0.0005 & 0.0126 & 0.0009 & 0.0103 & 0.41 & 0.0332 & 2.51 \\
\cmidrule(lr){2-9}
\multirow{5}{*}{MCTS} & Run 1 & 0.0284 & 0.2993 & 0.0452 & 0.5024 & 5.93 & 0.3546 & -26.50 \\
& Run 2 & 0.0277 & 0.3618 & 0.0398 & 0.5337 & 5.71 & 0.3253 & -31.79 \\
& Run 3 & 0.0294 & 0.3204 & 0.0384 & 0.5173 & 5.50 & 0.3288 & -35.40 \\
\cmidrule(lr){2-9}
& Mean & 0.0285 & 0.3272 & 0.0411 & 0.5178 & 5.71 & 0.3362 & -31.23 \\
& Std. & 0.0007 & 0.0260 & 0.0029 & 0.0128 & 0.18 & 0.0131 & 3.65 \\
\cmidrule(lr){2-9}
\multirow{5}{*}{Graph} & Run 1 & 0.0312 & 0.3318 & 0.0485 & 0.5449 & 8.87 & 0.5463 & -24.68 \\
& Run 2 & 0.0331 & 0.3370 & 0.0512 & 0.5714 & 8.57 & 0.5320 & -24.28 \\
& Run 3 & 0.0336 & 0.3712 & 0.0506 & 0.5935 & 7.92 & 0.5412 & -26.33 \\
\cmidrule(lr){2-9}
& Mean & 0.0326 & 0.3467 & 0.0501 & 0.5699 & 8.45 & 0.5398 & -25.10 \\
& Std. & 0.0010 & 0.0175 & 0.0012 & 0.0199 & 0.40 & 0.0059 & 0.89 \\
\bottomrule
\end{tabular*}
\end{table}

Using the three independent runs, CosmosAlpha-Graph has the highest mean in
all seven metrics. Its mean IC, ICIR, RankIC, and RankICIR are
$0.0326$, $0.3467$, $0.0501$, and $0.5699$, respectively, compared with
$0.0285$, $0.3272$, $0.0411$, and $0.5178$ for MCTS and
$0.0235$, $0.2750$, $0.0364$, and $0.4227$ for Chain. The same ordering holds
for the portfolio metrics: Graph reaches a mean ARR of $8.45\%$, IR of
$0.5398$, and signed MDD of $-25.10\%$, outperforming MCTS and Chain on the
mean risk--return profile. Run-to-run dispersion is metric-dependent. Graph's
ARR standard deviation is $0.40$, between MCTS ($0.18$) and Chain ($0.41$),
while its IR and MDD standard deviations ($0.0059$ and $0.89$) are lower than
those of MCTS ($0.0131$ and $3.65$) and Chain ($0.0332$ and $2.51$). Although
Graph is not the least variable method on every predictive metric, it leads
every metric on average, indicating that the overall advantage is not caused
by a single favorable run.

\clearpage
\section{Additional Controlled and Robustness Experiments}
\label{app:additional-controls}

This appendix reports supplemental CSI 300 experiments separately from the
main tables. Each study states its controls and uses the common downstream
metric definitions in Appendix~\ref{app:metrics}.

\subsection{Matched Generation and Token Budgets}
\label{app:matched-budget}

\paragraph{Generation control.}
Graph, MCTS, and Chain start from the same two evaluated expressions:
\begin{lstlisting}[style=prompt]
mul(mul(sub(close, low), inv(sub(high, low))), ts_rank(volume, 20))
rank(mul(sub(div(close, vwap), 1), div(turn, div(ts_sum(turn, 20), 20))))
\end{lstlisting}
All three use the following FactorDesign prompt verbatim. Graph and MCTS also
share Hypothesis/Design parent-information fields: the expression, raw RankIC
mean and standard deviation, and aligned Top-20 return, with common fields,
operators, and constraints. We denote the shared-generation Graph and MCTS
variants by Graph-G0 and MCTS-G0, respectively, with architecture-specific
archive-gap, neighborhood, and dimension/portrait/ancestor-Lesson guidance
omitted. Each controller retains its selection and state updates;
Chain-native retains its history/Feedback workflow, whose costs are included.

\begin{lstlisting}[style=prompt]
SYSTEM
You are the CosmosAlpha Factor Design Agent. Produce exactly one concise, testable daily equity-factor design. Use only supplied raw fields, registered operators and supported windows. Do not write DSL or executable code. When a parent expression is supplied, design a useful refinement grounded in the hypothesis and the parent raw quality metrics. Consider predictive strength, stability and top-20 return together. Return only the exact JSON object requested.

USER
Factor design request JSON:
${request_json}

Use `request_json.universe` as the research stock-pool scope. Do not replace it with another universe or use current or future constituent information.

Runtime capability view JSON:
${prompt_view_json}

Return a designs array containing exactly one item. Each design must contain exactly: name, description, mechanism_mapping, signal_definition, variables (objects with name and role), formulation, window_choices (objects with window and rationale), expected_direction, failure_conditions. Every variables[].name must exactly match one field name in Runtime capability view JSON.catalog.fields; use only its raw catalog name and do not list intermediate variables. variables[].role should describe the field's input role. Intermediate quantities belong only in formulation or signal_definition. Every window_choices[].window must be one of the supported_windows shown by the catalog. formulation must be mathematical or natural-language construction guidance, not DSL.
\end{lstlisting}

\paragraph{Budget and evaluation.}
All methods use DeepSeek-V4-Flash at temperature 0, common stage limits
(32,000 input Tokens; 20,000 output for expansion FactorDesign and 12,000 for
initialization and other stages), the same data split, validation/repair,
and LightGBM protocol. The 4,000,000 input-plus-output Token cap covers every
request, including failures, retries, repair, and Feedback. Missing provider
usage is charged at its reserved upper bound. Search stops when the next
request cannot fit; final factor counts are unconstrained. These runs are
separate from Table~\ref{tab:main-csi300}.

\begin{table}[!htbp]
\caption{CSI 300 test performance under shared generation conditions and a 4M-Token cap. G0 denotes the shared-generation control with architecture-specific guidance omitted. ARR and signed MDD are percentages; higher is better for all metrics. Bold marks the best value.}
\label{tab:matched-budget-performance}
\centering
\footnotesize
\setlength{\tabcolsep}{4pt}
\renewcommand{\arraystretch}{1.15}
\begin{tabular*}{\linewidth}{@{\extracolsep{\fill}}lrrrrrrr@{}}
\toprule
\multirow{2}{*}{Method} & \multicolumn{4}{c}{Predictive} & \multicolumn{3}{c}{Portfolio} \\
\cmidrule(lr){2-5} \cmidrule(lr){6-8}
& IC & ICIR & RankIC & RankICIR & ARR & IR & MDD \\
\midrule
Graph-G0 & \bestresult{0.0341} & \bestresult{0.3817} & \bestresult{0.0527} & \bestresult{0.6013} & \bestresult{8.29} & \bestresult{0.5823} & \bestresult{-23.92} \\
MCTS-G0 & 0.0293 & 0.3399 & 0.0444 & 0.5348 & 6.13 & 0.4494 & -25.20 \\
Chain-native & 0.0214 & 0.2301 & 0.0346 & 0.3994 & 3.08 & 0.1709 & -34.09 \\
\bottomrule
\end{tabular*}
\end{table}

\begin{table}[!htbp]
\caption{Search accounting for the 4M runs. Expansion cycles and controller steps exclude initialization; evaluations include it. Controller steps are parent attempts for Graph, tree iterations for MCTS, and serial cycles for Chain. Counts at different stages can overlap.}
\label{tab:matched-budget-search}
\centering
\footnotesize
\setlength{\tabcolsep}{3pt}
\renewcommand{\arraystretch}{1.1}
\begin{tabular}{@{}>{\raggedright\arraybackslash}p{0.29\linewidth}*{3}{>{\raggedright\arraybackslash}p{0.22\linewidth}}@{}}
\toprule
Measure & Graph-G0 & MCTS-G0 & Chain-native \\
\midrule
Structure / records & 173 graph nodes; archive 18 & 227 node occurrences (220 unique); AlphaZoo 18 & 100 rows \\
Evaluated factors & 176 & 225 & 95 \\
Expansion cycles & 218 & 260 & 113 \\
Controller steps & 214 parent attempts & 102 tree iterations & 113 serial cycles \\
Evaluations: events / unique & 176 / 176 & 226 / 225 & 96 / 95 \\
Unsuccessful outcomes & 43 recorded attempts + 4 cycles without attempt records & 34 distinct attempts & 14 cycles without factor records + 5 failed factors \\
Duplicate proposals / evaluations & 1 / 0 & 4 / 1 & 1 / 1 \\
\bottomrule
\end{tabular}
\end{table}

Graph excludes three low-date-coverage factors, giving $176-3=173$ nodes.
MCTS excludes five and retains seven repeated-expression occurrences, giving
$225-5+7=227$ nodes (220 unique expressions). Chain has 100 rows and
95 evaluated factors.

\begin{table}[!htbp]
\caption{LLM and Token accounting for the 4M runs. Calls include retries and repair; charged total equals reported usage plus the conservative missing-usage charge.}
\label{tab:matched-budget-tokens}
\centering
\footnotesize
\setlength{\tabcolsep}{3pt}
\renewcommand{\arraystretch}{1.1}
\begin{tabular}{@{}lrrr@{}}
\toprule
Measure & Graph-G0 & MCTS-G0 & Chain-native \\
\midrule
Calls: success / failure / parse failure & 683: 654 / 20 / 9 & 800: 780 / 6 / 14 & 468: 455 / 12 / 1 \\
Repair calls & 10 & 3 & 2 \\
Reported input & 1,993,999 & 2,282,457 & 2,734,050 \\
Reported output & 1,685,136 & 1,484,341 & 1,224,876 \\
Reported total & 3,679,135 & 3,766,798 & 3,958,926 \\
Missing-usage requests / charge & 6 / 272,000 & 4 / 184,000 & 0 / 0 \\
Charged total & 3,951,135 & 3,950,798 & 3,958,926 \\
Remaining budget & 48,865 & 49,202 & 41,074 \\
\bottomrule
\end{tabular}
\end{table}

Graph leads both alternatives on all seven test metrics, using fewer
factor evaluations and LLM calls than MCTS. This single-run control supports
the advantage of the implemented Graph controller under matched generation
conditions and Token budget. At the main experiments' approximately 150-factor
scale, estimated input-plus-output usage is 3.98M for Graph, 3.24M for MCTS,
and 7.84M for Chain, including conservative missing-usage charges.

\subsection{Normalization Sensitivity}
\label{app:normalization-sensitivity}

\paragraph{Saturation and frontier credit.}
The Quality Pareto Archive uses the raw objectives; normalization affects
SelfQuality and HV rewards. Across the three Graph runs in
Appendix~\ref{app:run-stability}, 467 factors provide the clipping denominator
in Table~\ref{tab:normalization-clipping}. The remaining 397 factors (85.01\%)
are interior to all three intervals. Comparisons between interior factors
retain the raw Pareto ordering; boundary contact alone does not establish a
changed comparison.

\begin{table}[!htbp]
\caption{Coordinates touching either clipping boundary in the three Graph runs.}
\label{tab:normalization-clipping}
\centering
\footnotesize
\setlength{\tabcolsep}{3pt}
\renewcommand{\arraystretch}{1.1}
\begin{tabular}{@{}lrr@{}}
\toprule
Coordinate at 0 or 1 & Count / factors & Percentage \\
\midrule
Absolute RankIC & 23/467 & 4.93\% \\
Reversed RankIC standard deviation & 14/467 & 3.00\% \\
Aligned Top-20 return & 56/467 & 11.99\% \\
At least one coordinate & 70/467 & 14.99\% \\
\bottomrule
\end{tabular}
\end{table}

In the same runs, 106 new factors improved the raw frontier. Birth gains
are evaluated independently against each frozen pre-round archive. We
recompute these same events using intervals 1.5 times as wide:
$[0,0.075]$, $[0.04,0.10]$, and $[0,0.0012]$ for RankIC, Std, and return;
twice as wide: $[0,0.10]$, $[0.03,0.11]$, and $[0,0.0016]$; and a smooth map
of the unclipped affine coordinates:
\begin{equation}
T(z)=\frac{1}{2}+\frac{\arctan(\pi(z-1/2))}{\pi}.
\label{eq:smooth-quality-normalization}
\end{equation}
For Std, set $z_{\mathrm{std}}=(\sigma(\mathrm{RankIC})-0.05)/(0.09-0.05)$
and use $1-T(z_{\mathrm{std}})$. This strictly monotone map has no hard
clipping. These comparisons hold the logged frontier events fixed.

\begin{table}[!htbp]
\caption{Zero normalized HV gains among the same 106 raw frontier improvements.}
\label{tab:normalization-zero-hv}
\centering
\footnotesize
\setlength{\tabcolsep}{3pt}
\renewcommand{\arraystretch}{1.1}
\begin{tabular}{@{}lrr@{}}
\toprule
Normalization & Zero HV / raw improvements & Percentage \\
\midrule
Original clipping & 8/106 & 7.55\% \\
Wider intervals, 1.5 times & 3/106 & 2.83\% \\
Wider intervals, 2 times & 0/106 & 0.00\% \\
Smooth atan map & 0/106 & 0.00\% \\
\bottomrule
\end{tabular}
\end{table}

Under the original normalization, 98/106 improvements (92.45\%) receive
positive HV credit. Wider bounds and smoothing reduce saturation in this
sample and offer reasonable alternatives to the empirical caps.

\paragraph{Fresh search with twice-wider intervals.}
We also rerun the full search with the twice-wider intervals and integrate the
resulting factor pool using the common CSI 300 protocol. This experiment
allows subsequent rewards, selections, and generated factors to change;
it is distinct from the fixed-history recalculation above.

\begin{table}[!htbp]
\caption{Original and twice-wider normalization after full search and LightGBM integration. ARR and signed MDD are percentages.}
\label{tab:normalization-fresh-search}
\centering
\footnotesize
\setlength{\tabcolsep}{4pt}
\renewcommand{\arraystretch}{1.15}
\begin{tabular*}{\linewidth}{@{\extracolsep{\fill}}lrrrrrrr@{}}
\toprule
\multirow{2}{*}{Setting} & \multicolumn{4}{c}{Predictive} & \multicolumn{3}{c}{Portfolio} \\
\cmidrule(lr){2-5} \cmidrule(lr){6-8}
& IC & ICIR & RankIC & RankICIR & ARR & IR & MDD \\
\midrule
Original Graph & 0.0324 & 0.3441 & 0.0494 & 0.5670 & 8.44 & 0.5332 & -25.07 \\
Twice-wider intervals & 0.0332 & 0.3313 & 0.0512 & 0.5538 & 7.86 & 0.5254 & -26.12 \\
\bottomrule
\end{tabular*}
\end{table}

Performance remains comparable to the original setting and within or close
to the ranges of the three-run stability study. This single supplemental
run supports the reasonableness of the original anchors while motivating
wider bounds or smoothing as further refinements.

\subsection{Separating Propagation and Diversity}
\label{app:relation-role-controls}

\paragraph{Controls.}
The original Structure and Correlation removals affect both Flow and
AverageDistance. For each relation, we add a Flow-only removal that leaves
AverageDistance intact and renormalizes the retained Flow weights, and a
Distance-only removal that retains the full Flow operator and weights while
omitting the named distance contribution. Full, each single-role removal, and
joint removal form a $2\times2$ comparison under the common CSI 300 ablation
protocol. Table~\ref{tab:relation-role-controls} includes the original Full and
joint-removal rows.

\begin{table}[!htbp]
\caption{Relation-role controls after LightGBM integration. The named role is removed; ARR and signed MDD are percentages. Bold marks the best value.}
\label{tab:relation-role-controls}
\centering
\footnotesize
\setlength{\tabcolsep}{4pt}
\renewcommand{\arraystretch}{1.15}
\begin{tabular*}{\linewidth}{@{\extracolsep{\fill}}lrrrrrrr@{}}
\toprule
\multirow{2}{*}{Removed relation / role} & \multicolumn{4}{c}{Predictive} & \multicolumn{3}{c}{Portfolio} \\
\cmidrule(lr){2-5} \cmidrule(lr){6-8}
& IC & ICIR & RankIC & RankICIR & ARR & IR & MDD \\
\midrule
Full Graph & \bestresult{0.0324} & \bestresult{0.3441} & \bestresult{0.0494} & \bestresult{0.5670} & \bestresult{8.44} & \bestresult{0.5332} & \bestresult{-25.07} \\
\addlinespace[2pt]
Structure: Flow only & 0.0303 & 0.3251 & 0.0464 & 0.5188 & 5.12 & 0.3145 & -29.01 \\
Structure: Distance only & 0.0311 & 0.3301 & 0.0469 & 0.5319 & 5.49 & 0.3283 & -28.19 \\
Structure: Flow + Distance & 0.0276 & 0.3022 & 0.0453 & 0.5289 & 4.37 & 0.2659 & -29.27 \\
\addlinespace[2pt]
Correlation: Flow only & 0.0303 & 0.3332 & 0.0476 & 0.5337 & 7.75 & 0.4715 & -26.22 \\
Correlation: Distance only & 0.0307 & 0.3316 & 0.0461 & 0.5118 & 7.52 & 0.4791 & -26.23 \\
Correlation: Flow + Distance & 0.0295 & 0.3290 & 0.0447 & 0.5304 & 7.28 & 0.4401 & -26.06 \\
\bottomrule
\end{tabular*}
\end{table}

Full leads all seven metrics. Each single-role removal exceeds its
joint-removal counterpart on most metrics, including ARR and IR. These
single-run controls support contributions through both propagation and
diversity, while retaining metric-specific exceptions.

\subsection{Rolling Robustness and Statistical Comparisons}
\label{app:rolling-robustness}

\paragraph{Rolling protocol.}
Five windows move all three temporal segments forward annually
(Table~\ref{tab:rolling-windows}). In every window, Graph, MCTS, and Chain
separately rerun the full LLM factor-mining process using only that window's
training data. Expressions are then fixed for validation and testing.
LightGBM fits on the same training segment with shared fixed hyperparameters;
the subsequent validation segment selects the boosting iteration. Alpha158
retains its predefined expressions and its model is retrained for each window.
These supplemental runs differ from those in Table~\ref{tab:main-csi300}.

\begin{table}[!htbp]
\caption{Rolling CSI 300 windows. Mining and model fitting use the same training segment; each test segment is a full calendar year.}
\label{tab:rolling-windows}
\centering
\footnotesize
\setlength{\tabcolsep}{3pt}
\renewcommand{\arraystretch}{1.1}
\begin{tabular}{@{}llll@{}}
\toprule
Window & Mining / training & Validation & Test \\
\midrule
2021 & Apr 2010--May 2019 & Jun 2019--Dec 2020 & 2021 \\
2022 & Apr 2011--May 2020 & Jun 2020--Dec 2021 & 2022 \\
2023 & Apr 2012--May 2021 & Jun 2021--Dec 2022 & 2023 \\
2024 & Apr 2013--May 2022 & Jun 2022--Dec 2023 & 2024 \\
2025 & Apr 2014--May 2023 & Jun 2023--Dec 2024 & 2025 \\
\bottomrule
\end{tabular}
\end{table}

\begin{table}[!htbp]
\caption{Annual out-of-sample RankIC after rolling factor mining and model fitting. Bold marks the highest value in each year.}
\label{tab:rolling-rankic}
\centering
\footnotesize
\setlength{\tabcolsep}{3pt}
\renewcommand{\arraystretch}{1.1}
\begin{tabular}{@{}lrrrr@{}}
\toprule
Test year & Alpha158 & Chain & MCTS & Graph \\
\midrule
2021 & 0.0560 & 0.0515 & 0.0562 & \bestresult{0.0573} \\
2022 & 0.0543 & 0.0437 & 0.0487 & \bestresult{0.0562} \\
2023 & 0.0461 & 0.0258 & 0.0323 & \bestresult{0.0481} \\
2024 & \bestresult{0.0465} & 0.0284 & 0.0378 & 0.0445 \\
2025 & 0.0546 & 0.0508 & 0.0572 & \bestresult{0.0611} \\
\bottomrule
\end{tabular}
\end{table}

\begin{table}[!htbp]
\caption{Annual out-of-sample ARR before transaction costs, in percent. Bold marks the highest value in each year.}
\label{tab:rolling-arr}
\centering
\footnotesize
\setlength{\tabcolsep}{3pt}
\renewcommand{\arraystretch}{1.1}
\begin{tabular}{@{}lrrrr@{}}
\toprule
Test year & Alpha158 & Chain & MCTS & Graph \\
\midrule
2021 & 9.49 & 10.48 & \bestresult{11.47} & 9.56 \\
2022 & -11.83 & -12.65 & -12.57 & \bestresult{-8.49} \\
2023 & \bestresult{-0.02} & -6.66 & -8.16 & -2.61 \\
2024 & 20.56 & 21.80 & 20.76 & \bestresult{23.84} \\
2025 & 11.02 & 13.60 & 15.21 & \bestresult{16.66} \\
\bottomrule
\end{tabular}
\end{table}

Graph's annual RankIC exceeds MCTS and Chain in all five years and Alpha158
in four; its ARR exceeds each baseline in four years. It ranks first among
all four methods in RankIC in four years and ARR in three.

\paragraph{Paired uncertainty estimates.}
To account for temporal dependence, we use 5,000 paired stationary-bootstrap
resamples~\citep{politis1994stationary} within each test year, with mean block
length 20 trading days. Random-start blocks have geometric lengths and wrap
within the year; the same sampled dates are used across methods. Each replicate
aggregates all five years with fixed valid-trading-day weights. RankIC uses
common valid signal dates and ARR common valid portfolio dates. We report
Graph-minus-baseline mean RankIC differences and annualized return differences
under Appendix~\ref{app:portfolio-metrics}, with 2.5th/97.5th percentile intervals.
Mean block lengths 10 and 40 give the same qualitative conclusions. The
intervals condition on the fitted runs and quantify test-period uncertainty.

\begin{table}[!htbp]
\caption{Paired Graph-minus-baseline effects with estimated 95\% stationary-bootstrap confidence intervals. ARR differences are percentage points.}
\label{tab:rolling-bootstrap}
\centering
\footnotesize
\setlength{\tabcolsep}{3pt}
\renewcommand{\arraystretch}{1.1}
\begin{tabular}{@{}lrr@{}}
\toprule
Baseline & RankIC difference [95\% CI] & ARR difference [95\% CI] \\
\midrule
Alpha158 & +0.00194 [-0.00069, 0.00458] & +1.95 [-1.71, 5.66] \\
MCTS & +0.00700 [0.00447, 0.00953] & +2.45 [-1.96, 6.99] \\
Chain & +0.01340 [0.00987, 0.01685] & +2.48 [-2.68, 7.59] \\
\bottomrule
\end{tabular}
\end{table}

All six point estimates favor Graph. RankIC intervals against MCTS and Chain
exclude zero; those against Alpha158 and all ARR intervals include zero.
The rolling comparisons complement the three fixed-split independent Graph
runs in Appendix~\ref{app:run-stability}: the studies assess temporal and
stochastic stability, respectively.

We distinguish positive out-of-sample predictive performance from statistically
significant factor discoveries. The reported bootstrap confidence intervals
assess the robustness of performance differences between integrated factor pools,
without implying statistical significance for every individual factor.

\subsection{Turnover and Transaction Costs}
\label{app:transaction-costs}

\paragraph{Cost replay.}
For the CSI 300 cost replay over 2021--2025
(Table~\ref{tab:transaction-costs}), Graph and Alpha158 use the predictions
underlying Table~\ref{tab:main-csi300}, while MCTS and Chain use Run~1 from
Appendix~\ref{app:run-stability}. All four follow the common backtest protocol.
Symmetric proportional costs of 0/5/10/20 basis points per side
are charged on buys and sells as combined commission/slippage scenarios.
Strategy settings remain fixed while the account is replayed at each cost
level. ARR, IR, and signed MDD follow the definitions in
Appendix~\ref{app:portfolio-metrics}. Annual turnover is mean daily turnover
multiplied by 238, matching the annualization factor used for ARR and IR.
The uncharged CSI 300 price-index reference covers the same 1,212 dates.

\begin{table}[p]
\caption{CSI 300 transaction-cost sensitivity and the uncharged market reference. Cost/side is in bps; ARR, signed MDD, and daily turnover are percentages; annual turnover is in times per year. Turnover counts both buys and sells.}
\label{tab:transaction-costs}
\centering
\footnotesize
\setlength{\tabcolsep}{4pt}
\renewcommand{\arraystretch}{1.15}
\begin{tabular*}{\linewidth}{@{\extracolsep{\fill}}lrrrrrr@{}}
\toprule
\multirow{2}{*}{Factor set} & \multirow{2}{*}{Cost/side} & \multicolumn{3}{c}{Portfolio} & \multicolumn{2}{c}{Turnover} \\
\cmidrule(lr){3-5} \cmidrule(lr){6-7}
& & ARR & IR & MDD & Daily & Annual \\
\midrule
Graph & 0 & 8.44 & 0.5332 & -25.07 & 19.11 & 45.48 \\
Graph & 5 & 6.28 & 0.3897 & -26.54 & 19.09 & 45.43 \\
Graph & 10 & 4.01 & 0.2485 & -30.88 & 19.07 & 45.39 \\
Graph & 20 & -0.53 & -0.0330 & -46.00 & 19.03 & 45.30 \\
\addlinespace[2pt]
MCTS & 0 & 5.93 & 0.3546 & -26.50 & 19.96 & 47.51 \\
MCTS & 5 & 3.55 & 0.2126 & -32.75 & 19.94 & 47.47 \\
MCTS & 10 & 1.18 & 0.0707 & -40.12 & 19.93 & 47.43 \\
MCTS & 20 & -3.55 & -0.2124 & -57.35 & 19.90 & 47.36 \\
\addlinespace[2pt]
Alpha158 & 0 & 6.32 & 0.3945 & -29.73 & 19.93 & 47.44 \\
Alpha158 & 5 & 3.97 & 0.2480 & -32.58 & 19.91 & 47.37 \\
Alpha158 & 10 & 1.62 & 0.1012 & -37.67 & 19.88 & 47.32 \\
Alpha158 & 20 & -3.07 & -0.1919 & -53.68 & 19.82 & 47.18 \\
\addlinespace[2pt]
Chain & 0 & 3.49 & 0.1839 & -39.27 & 19.31 & 45.97 \\
Chain & 5 & 1.18 & 0.0623 & -44.54 & 19.30 & 45.95 \\
Chain & 10 & -1.12 & -0.0590 & -50.65 & 19.29 & 45.91 \\
Chain & 20 & -5.70 & -0.3002 & -67.88 & 19.26 & 45.84 \\
\addlinespace[2pt]
CSI 300 index & -- & -0.78 & -0.0445 & -56.00 & -- & -- \\
\bottomrule
\end{tabular*}
\end{table}

Graph has the lowest turnover and leads the factor baselines on ARR, IR,
and signed MDD at every cost level. At 10 bps per side, its net ARR is 4.01\%
with IR 0.2485. At 20 bps, all four net ARRs are negative, yet Graph still
exceeds the uncharged CSI 300 index in ARR ($-0.53\%$ versus $-0.78\%$),
IR ($-0.0330$ versus $-0.0445$), and MDD ($-46.00\%$ versus $-56.00\%$).

\end{document}